\documentclass[12pt]{iopart}
\usepackage{amssymb}

\begin{document}

\title{Classical states, quantum field measurement}
\author{Peter Morgan}
\address{Physics Department, Yale University, New Haven, CT 06520, USA.}
\ead{peter.w.morgan@yale.edu}

\begin{abstract}
Classical Koopman--von Neumann Hilbert spaces of states are constructed here by the action of classical random fields on a vacuum state in ways that support an action of the quantized electromagnetic field and of the $U(1)$--invariant observables of the quantized Dirac spinor field, allowing a manifestly Lorentz invariant classical understanding of the state spaces of the two field theories, generalizing the Quantum--Mechanics--Free Systems of Tsang\&Caves and Quantum Non-Demolition measurements.
The algebra of functions on a classical phase space is commutative but the algebra of classical observables associated with coordinate transformations is noncommutative, so that, for example, we can as much ask whether a classical state is an eigenstate of a rotation as we can in quantum mechanics and so that entangled states can be distinguished from mixed states, making classical random fields as \emph{weird} as quantum fields.\vspace{2ex}

\hspace*{-2em}\fbox{\begin{minipage}{0.85\textwidth}
This is the Accepted Manuscript version of an article accepted for publication in \emph{Physica Scripta} (published April 16th, 2019.)  IOP Publishing Ltd is not responsible for any errors or omissions in this version of the manuscript or any version derived from it.  The Version of Record is available online at \textsf{https://doi.org/10.1088/1402-4896/ab0c53}.
\end{minipage}}
\end{abstract}

\

\maketitle

\newcommand\Intd{{\mathrm{d}}}
\newcommand\Half{{\scriptstyle\frac{\scriptstyle 1}{\scriptstyle 2}}}
\renewcommand\rmi{{\mathrm{i}}}
\renewcommand\rme{{\mathrm{e}}}
\newcommand\VEV[1]{{\left<0\right|#1\left|0\right>}}
\newcommand\Vacuum{{|0\rangle}}
\newcommand\VEVx[1]{{\left<{\scriptstyle\mathcal{F}}\right|#1\left|{\scriptstyle\mathcal{F}}\right>}}
\newcommand\VEVxL[1]{{\left<{\scriptstyle\mathcal{F}}\right|#1}}
\newcommand\VEVxR[1]{{#1\left|{\scriptstyle\mathcal{F}}\right>}}
\newcommand\VEVchi[1]{{\left<\mathbf{\circledcirc}\right|#1\left|\mathbf{\circledcirc}\right>}}
\newcommand\VEVchiR[1]{{#1\left|\mathbf{\circledcirc}\right>}}

\newcommand{\sfF}{{\mathsf{F}}}
\newcommand{\sff}{{\mathsf{f}}}
\newcommand{\sfg}{{\mathsf{g}}}
\newcommand{\sfb}{{\mathfrak{b}}}
\newcommand{\sfd}{{\mathfrak{d}}}
\newcommand{\sfU}{{\mathsf{U}}}
\newcommand{\sfV}{{\mathsf{V}}}
\newcommand{\sfa}{{\mathsf{a}}}
\newcommand{\Chi}{{\mathsf{X}}}
\newcommand{\bbFF}{{\hat\mathbb{F}}}
\newcommand{\iS}{{\rmi\hspace{-0.1em}\mathsf{S}}}
\newcommand{\Ej}{{\mathsf{j}}}

\section{Introduction}
If we are to understand the relationship between classical and quantum mechanics better than through the somewhat ill--defined process of quantization, one way to do so is to construct a Hilbert space formalism for classical mechanics, as Koopman did in 1931\cite{Koopman}, and then to construct and examine in detail a very different relationship between classical and quantum mechanics, of isomorphisms between Hilbert spaces and operator algebras.
We can call this a \emph{unification} of classical and quantum mechanics, just as Dirac constructed a unification of Schr\"odinger's differential equations and Heisenberg's matrix algebras as Hilbert space formalisms, even though we will find that there are significant differences: there are isomorphisms only for \emph{some} quantum field models (including, however, in particular, quantized electromagnetism).
Such isomorphisms between classical and quantum Hilbert space formalisms make consequences of the Hilbert space structure, such as the violation of Bell inequalities, Gleason's theorem, the Kochen-Specker paradox, and the rest, all not as definitive as they have been held to be: quantum is not distinguished from classical by Hilbert space structure and a noncommutative algebra of observables.
Discussions of when classical models can be effectively used as physical models will in future have to depend on more subtle distinctions.

It was shown in \cite{MorganEPL} that the Hilbert space of states over the complex Klein-Gordon quantum field is isomorphic to the Hilbert space of states over a real Klein-Gordon random field. {\small[A similar construction can be found in \cite[\S 5]{Cohn}.]}
In \cite{MorganEMunpublished} this was extended to show that the Hilbert space of states over the quantized electromagnetic field is isomorphic to the Hilbert space of states over an electromagnetic random field.
{\small[This construction generalizes the commutation of observables at time--like separation introduced as Quantum--Mechanics--Free Systems in quantum optics in \cite{TsangCaves} to commutation at arbitrary space-like or time--like separation, and a construction similar to this can be found in \cite{Ghose}.]}
The construction for the electromagnetic field will be reproduced and extended here, in Section~\ref{EMfield}, followed by a construction of an embedding of the algebra of global-$U(1)$ invariant observables of the quantized Dirac spinor field into an algebra of Dirac spinor random field operators in Section~\ref{QDirac}.
{\small[The construction given here for the Dirac spinor field has not previously appeared.]}

For the quantized electromagnetic field $\hat\sfF_{\mu\nu}(x)$, we can construct a commuting algebra $\mathcal{R}$ that is generated by a bivector--valued random field $\bbFF_{\mu\nu}(x)$ for $x$ anywhere in all of Minkowski space, not just, as one might first expect, on a space-like hyperplane.
The Hilbert space of the quantized electromagnetic field is generated by the action of $\bbFF_{\mu\nu}(x)$ on the vacuum state, then self-adjoint operators that act on the Hilbert space but do not commute with $\mathcal{R}$ generate transformations that in classical physics are generated by the Poisson bracket.

For the quantized Dirac spinor field, a Lie algebra $\mathcal{D}$ of global-U(1) invariant observables can be constructed as a subalgebra of a bosonic raising and lowering algebra, $\mathcal{D}\subset\mathcal{B}$, and the usual vacuum state over $\mathcal{D}$ can be extended (here, trivially) to act over $\mathcal{B}$, which contains a commutative subalgebra that corresponds to a random field on Minkowski space that is enough to construct the Hilbert space of states over $\mathcal{D}$.

\ref{CMtoQM} shows that we can for ordinary Classical Mechanics construct a Hilbert space over $\mathbb{R}$ by a Koopman--von Neumann approach\cite{Koopman,vonNeumann} (see \ref{CMtoQM} for recent references); the use of fourier transforms in field theory, however, provides a natural complex structure, allowing the construction of Hilbert spaces over $\mathbb{C}$, so that, paradoxically, random fields are somewhat closer to quantum fields than ordinary classical mechanics is to quantum mechanics.
{\small[Some readers will feel that \ref{CMtoQM} should be in the main text of the paper, before Section~\ref{EMfield}, because indeed \ref{CMtoQM} motivates and informs Section~\ref{EMfield}, however the mathematics in the main text does not at all depend on \ref{CMtoQM} (which, to keep a relative simplicity, does not follow the manifest Lorentz covariance of the main text).]}

Hilbert spaces for random fields that are isomorphic to Hilbert spaces for quantum fields will be constructed here only for free field cases.
Insofar as interacting QFT reduces to S-matrices\cite{Blum} ---which map from in-- to out--free field Hilbert spaces---, the same S-matrix works equally as well between random field Hilbert spaces that are isomorphic to those quantum field Hilbert spaces.
It's not {\sl necessary} to construct an interacting dynamics for the random field case insofar as we already have a successful interacting dynamics for the quantum field case, after regularization and renormalization, although hopefully having a random field construction also available may suggest new avenues.

The interpretation suggested here is that we can consider the states of free field quantum electrodynamics to be classical, for which the classical dynamics is a group of canonical transformations acting on the states that is generated by a classical Hamiltonian, with a parallel unitary quantum dynamics acting on the observables that is generated by a quantized Hamiltonian operator.
In the classical case, the action of the classical Hamiltonian requires the use of the Poisson bracket to construct the Liouville operator, whereas the quantized Hamiltonian has a direct adjoint action on observables.
The construction in the main text, however, will preserve manifest Lorentz and translation covariance throughout, implicitly specifying a Poincar\'e invariant stochastic dynamics.

There is also a subsidiary intention here to interpret quantum field theory as a stochastic signal analysis formalism, which in some empiricist sense quantum field theory has to be because experiments induce electrical and optical signals in cables, which are then statistically analyzed in hardware and software (for a variety of approaches to stochastic signal analysis, see \cite{BaezBiamonte,MumfordDesolneux,Rozanov}).
A quantum field operator (i) allows us to modulate the vacuum state, and (ii) allows us to make local measurements of those modulated states, so it is often appropriate to call the test functions that parameterize these operations (taken from a Schwartz space of functions that are smooth both in real space and in wave number space) either (i) ``modulation functions'' or (ii) ``window functions'' (alternatively, ``sampling functions'') depending on how they are used.
Insofar as general relativity can also be interpreted in terms of signals between places in space--time, a signal analysis interpretation for quantum field theory introduces more possibilities for unifying quantum theory with general relativity.

The existence of such constructions does not mean that this is the way the world is.
In particular, the constructions here are substantially nonunique, with there being many ways to construct the same system of states over the algebra of quantum field observables, so that it is absolutely necessary to be skeptical about any given model just as we are about Maxwell's vortices or about Bohmian trajectories.
Nonetheless, the constructions here \emph{are} manifestly Lorentz covariant, so they justify some further investigation of what advantages there might be in developing the classical dynamics of the random field states instead of developing the unitary dynamics of the quantum field states and observables.
In any case, note that we are determinedly working with Hilbert spaces, for which noncommuting observables are natural, even if the way in which states are constructed can be (but does not have to be) construed as more--or--less classical.

The notation used here may be offputtingly novel except for mathematical physicists, however it has a moderately principled motivation as an intrinsic vector formalism in the test function space (which for free fields are equipped with a pre--inner product).

\section{The electromagnetic field}\label{EMfield}
We can construct the quantized electromagnetic field
   $\hat\sfF_{\sff}=a^{\ }_{\sff^*}+a_{\sff}^\dagger$,
$\hat\sfF^\dagger_{\sff}=\hat\sfF_{\sff^*}$, for bivector test functions $\sff_{\mu\nu}(x)$, using a set of raising and lowering operators for which
$$\VEV{\hat\sfF^\dagger_{\sff}\hat\sfF_{\sfg}}=[a^{\ }_{\sff},a_\sfg^\dagger]=(\sff,\sfg)_+,
   \qquad[a^{\ }_{\sff},a_\sfg^{\ }]=0,\qquad\langle 0|a_\sff^\dagger=0=a^{\ }_{\sff}|0\rangle,
$$
where
\begin{eqnarray*}
 (\sff,\sfg)_\pm&=&
   -\hbar\!\int\!\widetilde{\delta \sff}^*(k){\cdot}\widetilde{\delta \sfg}(k)
                         2\pi\delta(k{\cdot}k)\theta(\pm k_0)\frac{\Intd^4k}{(2\pi)^4}\\
 &=&
   -\hbar\!\int\!k^\alpha{{\tilde\sff}_{\alpha\mu}\strut}^{\hspace{-1.6ex}*\hspace{0.6ex}}(k)g^{\mu\nu}
                           k^\beta\tilde\sfg_{\beta\nu}(k)
                         2\pi\delta(k{\cdot}k)\theta(\pm k_0)\frac{\Intd^4k}{(2\pi)^4},
\end{eqnarray*}
which are positive semi--definite sesquilinear forms on the vector space of test functions because $k^\alpha\tilde\sff_{\alpha\mu}^*(k)$ and $k^\beta\tilde\sfg_{\beta\nu}(k)$ are space--like 4--vectors that are orthogonal to the light--like 4--vector $k$ (the form of this \emph{pre--inner product} is derived in \cite[Eq. (3.27)]{MenikoffSharp}, for example).
The commutation relation $[\hat\sfF_{\sff},\hat\sfF_{\sfg}]=(\sff^*,\sfg){-}(\sfg^*,\sff)$ is zero if the supports of $\sff$ and $\sfg$ are space--like separated.\newline
\framebox[\columnwidth]{\hspace{0.005\columnwidth}\parbox{0.97\columnwidth}{\small\vspace*{0.3ex}\raggedright
All of the above can be derived by constructing $\hat\sfF_{\sff}$ as $\hat\sfF_{\sff}=\int\sff^{\mu\nu}\!(x)\hat\sfF_{\mu\nu}(x)\Intd^4x$, but a principal aim of the notation is to work intrinsically in test function space, and manifestly Lorentz covariantly, as far as possible.
We can revert to a point--like quantum field at a point $y$ by taking an improper test function $\sff^{\mu\nu}\!(x)$ that is a multiple of a Dirac delta function $\delta(x-y)$ or to the fourier transform of the quantum field at wave--number $k$ by taking an improper test function that is a multiple of $\rme^{\rmi k{\cdot}x}$, however careful models typically use less singular test functions that include details such as line widths or pulse durations; indeed experiments may be intended to produce states, such as Bessel beams, that require very carefully shaped test functions as models.
It is helpful that all geometrical information is isolated in the pre--inner product, so that for the real Klein--Gordon quantum field an intrinsic vector in a test function space formalism has exactly the same presentation, except only that the pre--inner product for scalar test functions $f$ and $g$ is different,\vspace{-1.5ex}
$$(f,g)_+=\hbar\!\int \tilde f^*(k)\tilde g(k)2\pi\delta(k{\cdot}k-m^2)\theta(k_0)\frac{\Intd^4k}{(2\pi)^4},
       \vspace{-1ex}$$
and a comparable presentation is possible for the complex Klein--Gordon quantum field, given in \ref{complexQKG}.
It is also helpful that it is easy to make a connection to quantum optics or to elementary discussions of quantum mechanics if we simply abbreviate $\hat\sfF_{\sff_1}$, $\hat\sfF_{\sff_2}$, ..., $\hat\sfF_{\sff_n}$, as $\hat\sfF_1$, $\hat\sfF_2$, ..., $\hat\sfF_n$, and similarly for raising and lowering operators.
The usually constructed quantized electromagnetic field, which supports an irreducible representation of the Poincar\'e group, follows as an inductive limit, taking the induction to be over a Schwartz space of smooth test functions on Minkowski space.
}}\vspace{0.5ex}

We introduce projection to left and right helicity, $\sff\mapsto \Half(1\pm\rmi\star)\sff$, using the Hodge dual $\star$,
$[\star\sff]_{\mu\nu}=\Half{\varepsilon_{\mu\nu}}^{\alpha\beta}\sff_{\alpha\beta}$, $\star\hspace{-0.55ex}\star\!\sff=-\sff$ (when acting on 2--forms), for which we find
$(\star\sff,\sfg)_\pm=-(\sff,\star\sfg)_\pm$ and hence
$(\Half(1\pm\rmi\star)\sff,\sfg)_\pm=(\sff,\Half(1\pm\rmi\star)\sfg)_\pm$,
so that we can construct raising and lowering operators for independent left and right helicity components of the quantized electromagnetic field,
\begin{eqnarray*}
 && \hat\mathsf{l}_{\sff}=a_{\Half(1+\rmi\star)\sff},\qquad
      \hat\mathsf{r}_{\sff}=a_{\Half(1-\rmi\star)\sff},\qquad
       [\hat\mathsf{l}_{\sff},\hat\mathsf{r}_{\sfg}^\dagger]=0,\\
 && [\hat\mathsf{l}_{\sff},\hat\mathsf{l}_{\sfg}^\dagger]=(\sff,\Half(1+\rmi\star)\sfg)_+,\ \quad
        [\hat\mathsf{r}_{\sff},\hat\mathsf{r}_{\sfg}^\dagger]=(\sff,\Half(1-\rmi\star)\sfg)_+,\\
 && \hat\sfF_{\sff}=\hat\mathsf{l}_{\sff^*}+\hat\mathsf{l}^\dagger_\sff
                                 +\hat\mathsf{r}_{\sff^{*}}+\hat\mathsf{r}^\dagger_{\sff}.
\end{eqnarray*}
Using the left and right helicity components and adopting the reversal introduced in \cite{MorganEPL},  $\sff^-(x)=\sff(-x)$, $\widetilde{\sff^-}(k)=\tilde\sff(-k)$, we construct one of the possible alternative field objects,  $\bbFF_{\sff}$, which will prove to be classical in the sense that the commutator is zero for all test functions, $[\bbFF_{\sff},\bbFF_{\sfg}]=0$,
\begin{eqnarray*}
  \bbFF_{\sff}&=&\hat\mathsf{l}_{\sff^*}+\hat\mathsf{l}^\dagger_\sff
                                 +\hat\mathsf{r}_{\sff^{-*}}+\hat\mathsf{r}^\dagger_{\sff^-},
        \hspace{6em} \bbFF_\sff^\dagger=\bbFF_{\sff^*},\\
    &=&a^{\ }_{\Half(1+\rmi\star)\sff^*+\Half(1-\rmi\star)\sff^{-*}}
         +a^\dagger_{\Half(1+\rmi\star)\sff+\Half(1-\rmi\star)\sff^-}
\end{eqnarray*}
so that, using the identities $(\sff^*,\sfg^*)_\pm=(\sfg,\sff)_\mp$ and $(\sff^-,\sfg^-)_\pm=(\sff,\sfg)_\mp$,
\begin{eqnarray*}\hspace{-5em}
  \VEV{\bbFF_{\sff}\bbFF_{\sfg}}
            =\VEV{\Bigl[\hat\mathsf{l}_{\sff^*}\hat\mathsf{l}^\dagger_\sfg
                            +\hat\mathsf{r}_{\sff^{-*}}\hat\mathsf{r}^\dagger_{\sfg^-}\Bigr]}
  &=&
            \Bigl(\sff^*,\Half(1{+}\rmi\star)\sfg\Bigr)_{\!+}
          \!+\Bigl(\sff^{-*},\Half(1{-}\rmi\star)\sfg^-\Bigr)_{\!+}\\
  &=&
            \Bigl(\sff^*,\Half(1{+}\rmi\star)\sfg\Bigr)_{\!+}
          \!+\Bigl(\sfg^{-*},\Half(1{+}\rmi\star)\sff^-\Bigr)_{\!-}\\
  &=&
            \Bigl(\sff^*,\Half(1{+}\rmi\star)\sfg\Bigr)_{\!+}
          \!+\Bigl(\sfg^{*}\hspace{0.65em},\Half(1{+}\rmi\star)\sff\hspace{0.65em}\Bigr)_{\!+}
             =\VEV{\bbFF_{\sfg}\bbFF_{\sff}}
\end{eqnarray*}
is symmetric, and we can similarly show that $[\bbFF_{\sff},\bbFF_{\sfg}]=0$.
$\bbFF_{\sff}$ is a classical Gaussian observable with variance $\VEV{\bbFF_{\sff}\bbFF_{\sff}}$ whenever $\sff=\sff^*$; however $\bbFF_{\sff}$, as a normal operator\cite{Zurek,Roberts}, is an observable even when $\sff\not=\sff^*$ in the sense that all components of its real and imaginary parts are jointly measurable.
Note that both the random field and the quantum field are translation invariant even though the transformation between them is not, because $(\sff,\sfg)_\pm$ are both translation invariant.

For any state that we could construct by the action of a function of $\hat\sfF_{\sff_1},...,\hat\sfF_{\sff_n}$ on the vacuum vector $|0\rangle$, for some set of bivector test functions $\{\sff_i(x)\}$, we can construct the same state by the action of some function of $\bbFF_{\sff_1^\bullet},...,\bbFF_{\sff_n^\bullet}$, because the linear map $\sff\mapsto \sff^\bullet=\Half(1+\rmi\star)\sff+\Half(1-\rmi\star)\sff^-$ is an involution, $\sff^{\bullet\bullet}=\sff$.
Consequently, we can, if we wish, say that the states are classical even if we continue to say that the measurements are nontrivially quantum mechanical.
The algebras that are generated by $\hat\sfF_\sff$ and by $\bbFF_{\sff}$ are both subalgebras of the commonplace raising and lowering algebra that is generated by $a_\sff^{\,}$ and $a_\sff^\dagger$,
$$
  \hat\sfF_\sff=a^{\ }_{\sff^*}+a_{\sff}^\dagger,\qquad
  \bbFF_{\sff}=a^{\ }_{\sff^{*\bullet}}+a_{\sff^\bullet}^\dagger,
$$
which are different because $\sff^{*\bullet}{\not=}\sff^{\bullet*}$, so we can more--or--less say that the electromagnetic random field we have constructed here has been hiding in plain sight.
With this presentation of $\hat\sfF_\sff$ and of $\bbFF_{\sff}$ and using the vacuum state, the isomorphism of the two Hilbert spaces they generate can be presented as an equality of normal--ordered expressions,
$$
  a_{\sfg_1}^\dagger{\cdots}\;a_{\sfg_n}^\dagger|0\rangle
  ={:}\hat\sfF_{\sfg_1}{\cdots}\;\hat\sfF_{\sfg_n}{:}|0\rangle
  ={:}\bbFF_{\sfg_1^\bullet}{\cdots}\;\bbFF_{\sfg_n^\bullet}{:}|0\rangle,
$$
with the Hilbert space inner product determined by the pre--inner product
$$\VEV{a_{\sff_n}{\cdots}\;a_{\sff_1}  a_{\sfg_1}^\dagger{\cdots}\;a_{\sfg_n}^\dagger}.
$$
We can equally take the opposite perspective, however, that the quantized electromagnetic field was always hiding in plain sight in the full algebra of observables of classical electromagnetism, when we apply the constructions of \ref{CMtoQM}.

When considering how committed we should be to either a quantum or a random field perspective, it is instructive to consider how flexibly committed quantum field theory is to microcausality.
In particular, quantum field models commonly introduce state transition probabilities as models of measurements, $\left|\langle S_1|S_2\rangle\right|^2=\langle S_1|S_2\rangle\langle S_2|S_1\rangle$, measurements of the projection--valued observable $|S_2\rangle\langle S_2|$ in the state $\langle S_1|\,{\cdot}\,|S_1\rangle$ (with both normalized); weighted sums of projection operators $|S_i\rangle\langle S_i|$ generate the space of normal operators on the Hilbert space of states.
The algebra of observables generated by $|S_i\rangle\langle S_i|$ can be generated by using the vacuum projection operator $|0\rangle\langle 0|$, and in general we can construct operators of the form
$$\hat\mathbf{O}(\underline{\sfg};\underline{\sff})=a_{\sfg_1}^\dagger{\cdots}\;a_{\sfg_n}^\dagger|0\rangle\langle 0|a_{\sff_n}{\cdots}\;a_{\sff_1}
$$
and sums of such operators, in which case we note that $|0\rangle\langle 0|$ and $\hat\mathbf{O}(\underline{\sfg};\underline{\sff})$ are essentially global operators,
$\bigl[|0\rangle\langle 0|,\hat\sfF_\sff\bigr]{\not=}0,$
$\bigl[\hat\mathbf{O}(\underline{\sfg};\underline{\sff}),\hat\sfF_\sff\bigr]{\not=}0\,\forall\sff$ (and, for the random field $\bbFF_\sff$, $\bigl[|0\rangle\langle 0|,\bbFF_\sff\bigr]{\not=}0,$
$\bigl[\hat\mathbf{O}(\underline{\sfg};\underline{\sff}),\bbFF_\sff\bigr]{\not=}0\,\,\forall\sff$), which do not satisfy microcausality.
Quantum field models commonly also discuss the measurement of number operators such as $\hat \mathsf{N}_{\sff}{=}a^\dagger_\sff a^{\ }_\sff$, which also do not satisfy microcausality.
If we allow the use of the vacuum projection operator for random field models, as indeed is appropriate for even moderately sophisticated classical signal analysis (where time--frequency analysis for non-stochastic signals is well--known to require Wigner and other quasi-distributions\cite{Cohen} because of the ubiquitous use of the fourier transform), the algebras of observables as well as the Hilbert spaces are identical.

For the quantized electromagnetic field, the noncommutative algebra of operators that is generated by $\hat\sfF_\sff$ \emph{and} the vacuum projection operator $|0\rangle\langle 0|$ is isomorphic to the algebra of operators that is generated by $\bbFF_\sff$ \emph{and} the vacuum projection operator $|0\rangle\langle 0|$, because the vacuum projection operator $|0\rangle\langle 0|$ does not commute with the commutative algebra generated by the $\bbFF_\sff$.
Including just this one projection operator $|0\rangle\langle 0|$ is enough to generate the same noncommutative algebra using the $\bbFF_\sff$ as is generated by $|0\rangle\langle 0|$ and using the $\hat\sfF_\sff$ because the Hilbert space is cyclic, generated by the action of the creation operators on the vacuum vector.
The algebras of operators that are generated by $\hat\sfF_\sff$ and by $\bbFF_\sff$ are both subalgebras of the algebra generated by the raising and lowering operators.

A significant difference between the random and quantum free field algebras is the \emph{irreducibility}\cite[p.101]{SW} (also referred to as \emph{completeness}\cite[\S II.1.2]{Haag}) of the algebra generated by the quantum field operators $\hat\sfF_\sff$, so that the quantum algebra can be considered preferable to the commutative algebras generated by $\bbFF_\sff$.
The Poisson or Peierls bracket, however, generates an algebra of transformations of the space of states or trajectories, so that if we include canonical transformations between different commutative subalgebras as an intrinsic part of classical physics then there is no difference between the two constructions, insofar as classical physics then does not give preference to any one maximally commutative subalgebra.
It is natural for classical physics that the vacuum projection operator \emph{must} be introduced to augment the algebra generated by the random field operators $\bbFF_\sff$ so that the transformations generated by the Poisson or Peierls bracket, which are arguably as important as part of the full classical theory as the observables, can be constructed.
We \emph{can} take the random field operators $\bbFF_\sff$ to be a possible way to generate probability densities over a set of \emph{beables} of a modal interpretation of the theory (somewhat like finding a Lorentz invariant and Hilbert space equivalent of de Broglie-Bohm trajectories), however to do so is to ignore the classical algebra of transformations, under which, as in quantum theory, we would not prioritize one maximal commutative subalgebra over another.
Classical mechanics ---when fully construed as including the Poisson or Peierls bracket, the algebra of transformations, and alternative maximal commutative subalgebras--- has always been as ``weird'' as quantum mechanics.
With the introduction of the vacuum projection operator, we must also note that a Koopman--von Neumann vacuum state over the full noncommutative algebra gives significantly more information than a Liouville--type state over the commutative algebra generated by the random field operators $\bbFF_\sff$ alone, particularly because a commutative algebra cannot distinguish mixtures from superpositions.

For any quantum field for which there is an involution on the test function space,
${}^\bullet:\mathcal{S}\rightarrow\mathcal{S}; f\mapsto f^\bullet, f^{\bullet\bullet}=f$ and for which the pre--inner product satisfies $(f^{*\bullet},g^\bullet)=(g^{*\bullet},f^\bullet)$, whereas in general $(f^*,g)\not=(g^*,f)$, we can construct a random field that is equivalent in the sense we have seen for the quantized electromagnetic field (the construction for the complex Klein-Gordon quantum field is given in \ref{complexQKG}).
This can be thought of as introducing a new type of reflection positivity\cite{Neeb}, which, however, preserves the 1+3--signature of space--time.

We have presented the construction above using raising and lowering operators.
Avoiding their use, we can present the quantum and random field structures for the electromagnetic field as
\newline\centerline{\begin{tabular}{|c|c|}
\hline{}&{}\vspace{-1ex}\\
$\displaystyle\hat\sfF_\sff^\dagger=\hat\sfF_{\sff^*}$ &
   $\displaystyle\bbFF_\sff^\dagger=\bbFF_{\sff^*}$\\
\quad$\displaystyle[\hat\sfF_{\sff},\hat\sfF_{\sfg}]=(\sff^*,\sfg)_+{-}(\sfg^*,\sff)_+$~~~&
   $\displaystyle[\bbFF_{\sff},\bbFF_{\sfg}]=0$\\
$\displaystyle\VEV{\hat\sfF_{\sff}\hat\sfF_{\sfg}}=(\sff^*,\sfg)_+$&
   \quad$\displaystyle\VEV{\bbFF_{\sff}\bbFF_{\sfg}}=(\sff^{*\bullet},\sfg^\bullet)_+$~~~\vspace{-1ex}\\
{}&{}\\
\hline\multicolumn{2}{|c|}{}\vspace{-1ex}\\
\multicolumn{2}{|c|}{\qquad$\displaystyle{:}\hat\sfF_{\sfg_1}{\cdots}\;\hat\sfF_{\sfg_n}{:}|0\rangle
  ={:}\bbFF_{\sfg_1^\bullet}{\cdots}\;\bbFF_{\sfg_n^\bullet}{:}|0\rangle$~~~}\vspace{-1.5ex}\\
\multicolumn{2}{|c|}{}\\
\hline
\end{tabular}}\vspace{1ex}\newline
with all other connected Wightman functions being zero.
The final equality, which assumes appropriately different definitions of normal--orderings for the two cases, identifies the Hilbert spaces of the two theories, defining an action of the algebra generated by the $\hat\sfF_{\sff}$ on the Hilbert space generated by $\bbFF_{\sff}$.
It is also worthwhile to present the construction above in a more unified Weyl--like coherent state formalism, using $\hat\mathsf{W}(\sff)=\rme^{\rmi\hat\sfF_\sff}$ and
$\hat\mathbb{W}(\sff)=\rme^{\rmi\bbFF_\sff}$,
\newline\centerline{\begin{tabular}{|c|c|}
\hline{}&{}\vspace{-1ex}\\
$\displaystyle\hat\mathsf{W}(\sff)^\dagger=\hat\mathsf{W}(-\sff^*)$ &
   $\displaystyle\hat\mathbb{W}(\sff)^\dagger=\hat\mathbb{W}(-\sff^*)$\\
\quad$\displaystyle\hat\mathsf{W}(\sff)\hat\mathsf{W}(\sfg)
          =\rme^{-\left[(\sff^*,\sfg)_+{-}(\sfg^*,\sff)_+\right]/2}\hat\mathsf{W}(\sff+\sfg)$\quad&
   $\displaystyle\displaystyle\hat\mathbb{W}(\sff)\hat\mathbb{W}(\sfg)
          =\hat\mathbb{W}(\sff+\sfg)$\\
$\displaystyle\VEV{\hat\mathsf{W}(\sff)}=\rme^{-(\sff^*,\sff)_+/2}$&
   \quad$\displaystyle\VEV{\hat\mathbb{W}(\sff)}
               =\rme^{-(\sff^{*\bullet},\sff^\bullet)_+/2}$\quad\vspace{-1ex}\\
{}&{}\\
\hline\multicolumn{2}{|c|}{}\vspace{-1ex}\\
\multicolumn{2}{|c|}{\hspace{6em}$\displaystyle
    \rme^{(\sff^*,\sff)_+/2}\hat\mathsf{W}(\sff)|0\rangle=
    \rme^{(\sff^{\bullet*\bullet},\sff)_+/2}\hat\mathbb{W}(\sff^\bullet)|0\rangle$
                           }\vspace{-1.5ex}\\
\multicolumn{2}{|c|}{}\\
\hline
\end{tabular}}\vspace{1ex}\newline
which can be seen to equate appropriately scaled coherent states.
The expression $\sff^{\bullet*\bullet}$ can be simplified to $\sff^{\bullet*\bullet}=\sff^{*-}=\sff^{-*}$:
\begin{eqnarray*}
  \widetilde{\sff^{\bullet*\bullet}}(k)&=&\Half(1+\rmi\star)\widetilde{\sff^{\bullet*}}(k)
                                                    +\Half(1-\rmi\star)\widetilde{\sff^{\bullet*}}(-k)\cr
     &=&\!\left[\Half(1-\rmi\star)\widetilde{\sff^{\bullet}}(-k)
                +\Half(1+\rmi\star)\widetilde{\sff^{\bullet}}(k)\right]^*\cr
     &=&\Big[\Half(1-\rmi\star)\left\{\Half(1+\rmi\star)\tilde\sff(-k)
                                               +\Half(1-\rmi\star)\tilde\sff(k)\right\}\cr
     &&\hspace{1em} +\Half(1+\rmi\star)\left\{\Half(1+\rmi\star)\tilde\sff(k)
                                                              +\Half(1-\rmi\star)\tilde\sff(-k)\right\}\Big]^*\cr
     &=&\left[\Half(1-\rmi\star)\tilde\sff(k)+\Half(1+\rmi\star)\tilde\sff(k)\right]^*\cr
     &=&\tilde\sff^*(k)=\widetilde{\sff^{*-}}(k)=\widetilde{\sff^{-*}}(k).
\end{eqnarray*}

So as not to delay the discussion of the quantized Dirac spinor field, a slightly complicated discussion of potentials for $\bbFF_{\sff}$ is presented in \ref{EMpotential}.

\section{The quantized Dirac spinor field}\label{QDirac}
First a terse presentation of the quantized Dirac spinor field: we can construct the field operators $\hat\psi^{\ }_{U}=d^{\ }_{U^c}+b_U^\dagger$
for a Dirac spinor test function $U$ using anticommuting raising and lowering operators for which
\begin{eqnarray*}
     &&\{b^{\ }_{U},b_V^\dagger\}=\{d^{\ }_{U},d_V^\dagger\}=(U,V)_+,\\
     &&  \{b^{\ }_{U},b^{\ }_V\}=\{b^{\ }_{U},d^{\ }_V\}=\{d^{\ }_{U},d^{\ }_V\}
                        =\{b^{\ }_{U},d^{\dagger}_V\}=0,\\
     &&  \langle 0|b_V^\dagger=\langle 0|d_V^\dagger=0=d_V^{\ }|0\rangle=b_V^{\ }|0\rangle
\end{eqnarray*}
where for the pre--inner product $(U,V)_+$ and its negative--frequency counterpart $(U,V)_-$, which is also positive semi--definite, we have
\begin{eqnarray*}(U,V)_\pm{=}\pm\hbar\!\!\int\!\overline{\tilde U(k)}(k{\cdot}\gamma+m)\tilde V(k)
                         \theta(\pm k_0)\Intd\mu_m(k);
\end{eqnarray*}
the measure $\Intd\mu_m(k)=2\pi\delta(k{\cdot}k-m^2)\frac{\Intd^4k}{(2\pi)^4}$ on the wave--number space is zero except on the forward and backward mass shells for mass $m$.
With this construction,
\begin{eqnarray*}
  &&\VEV{\hat\psi_U^\dagger\hat\psi^{\ }_{V}}=\VEV{b^{\ }_{U}b_V^\dagger}=(U,V)_+\\
  &&\VEV{\hat\psi^{\ }_{V}\hat\psi_U^\dagger}=\VEV{d^{\ }_{V^c}d_{U^c}^\dagger}=(V^c,U^c)_+=(U,V)_-\\
  &&\quad\{\hat\psi_U^\dagger,\hat\psi^{\ }_{V}\}=(U,V),\quad \{\hat\psi^{\ }_{U},\hat\psi^{\ }_{V}\}=0,
\end{eqnarray*}
where $(U,V) = (U,V)_++(U,V)_-$ [see below for a derivation of $(V^c,U^c)_\pm=(U,V)_\mp$].
The anti--commutator $\{\hat\psi_U^\dagger,\hat\psi^{\ }_{V}\}$ is zero if the supports of the test functions $U$ and $V$ are space--like separated.
Explicitly, the vital equalities $(U^c,V^c)_\pm=(V,U)_\mp$ depend on the identities $\overline{A^c}\gamma^\mu B^c\,{=}\,\overline{B}\gamma^\mu A$ and
$\overline{A^c}\,B^c\,{=}\,{-}\overline{B}\,A$,
because of which the arbitrary phase introduced by charge conjugation cancels,\vspace{-1ex}
\begin{eqnarray*}
(U^c,V^c)_\pm
                       &=&\pm\hbar\!\int\!\overline{\widetilde{U^c}(k)}
                            (k{\cdot}\gamma+m)\widetilde{V^c}(k)\theta(\pm k_0)\Intd\mu_m(k)\\
                       &=&\pm\hbar\!\int\!\overline{\left[\widetilde{V^c}(k)\right]^c}
                            \!(k{\cdot}\gamma-m)\left[\widetilde{U^c}(k)\right]^c\!\theta(\pm k_0)\Intd\mu_m(k)\\
                       &=&\pm\hbar\!\int\overline{\tilde V(-k)}
                            (k{\cdot}\gamma-m)\tilde U(-k)\theta(\pm k_0)\Intd\mu_m(k)\\
                       &=&\mp\hbar\!\int\overline{\tilde V(k)}
                            (k{\cdot}\gamma+m)\tilde U(k)\theta(\mp k_0)\Intd\mu_m(k)\\
 &=& (V,U)_\mp, \mbox{ so that, also, }(U^c,V^c)=(V,U).
\end{eqnarray*}
All of the above can be derived by setting $\hat\psi_U=\int\overline{U^c(x)}\hat\psi(x)\Intd^4x$ (with charge conjugation introduced to ensure that the inessential but usual convention for quantum fields is followed, that $\hat\psi_U$ is linear in the test function $U$), but, again, it is a principal aim to work intrinsically in test function space as far as possible.

To proceed, we note that it is generally understood that for an operator to be an observable of the quantized Dirac spinor free field formalism it has to be invariant under global-$U(1)$ transformations ($U(1)$--gauge transformations are considered briefly in Section~\ref{LocalU1}), for which the simplest case, $\hat\Phi^{\ }_U=\hat\psi_U^\dagger\hat\psi^{\ }_U=\hat\Phi^{\dagger}_U$, is a multiple of a projection operator,
$\hat\Phi^{\ 2}_U=\hat\psi_U^\dagger\hat\psi^{\ }_U\hat\psi_U^\dagger\hat\psi^{\ }_U
  =\hat\psi_U^\dagger\left((U,U)-\hat\psi_U^\dagger\hat\psi^{\ }_U\right)\hat\psi^{\ }_U
  =(U,U)\hat\Phi^{\ }_U$, so that $\frac{\hat\Phi^{\ }_U}{(U,U)}$ can be used to model yes/no measurement results, and is also local in that the commutator is
\begin{eqnarray*}
  [\hat\Phi^{\ }_U,\hat\Phi^{\ }_V]
    &=&[\hat\psi_U^\dagger\hat\psi^{\ }_U,\hat\psi_V^\dagger\hat\psi^{\ }_V]\\
    &=&\hat\psi_U^\dagger[\hat\psi^{\ }_U,\hat\psi_V^\dagger\hat\psi^{\ }_V]
          +[\hat\psi_U^\dagger,\hat\psi_V^\dagger\hat\psi^{\ }_V]\hat\psi^{\ }_U\\
    &=&(V,U)\hat\psi_U^\dagger\hat\psi^{\ }_V-(U,V)\hat\psi_V^\dagger\hat\psi^{\ }_U\\
    &=&\rmi\sqrt{(U,U)(V,V)}\Bigl[\hat\Phi^{\ }_{Y(V,U)}-\hat\Phi^{\ }_{Y(U,V)}\Bigl],\\
\mbox{where\qquad}&&\\
    Y(U,V)&=&U\sqrt{\frac{\sqrt{(U,V)(V,U)}}{2(U,U)}}+\ \rmi\frac{V(V,U)}{\sqrt{2(V,V)\sqrt{(U,V)(V,U)}}},
\end{eqnarray*}
so that $[\hat\Phi^{\ }_U,\hat\Phi^{\ }_V]$ is zero when the test functions $U$ and $V$ have space--like separated supports and so that the commutator $[\hat\Phi^{\ }_U,\hat\Phi^{\ }_V]$ is closed in the algebra generated by $\hat\Phi^{\ }_U$ (note that $Y(U,V)$ is defined only up to a phase, but the presentation here uses only $\sqrt{(U,V)(V,U)}$, avoiding the use of $\sqrt{(U,V)}$ alone or of $\sqrt{(V,U)}$).
It is sufficient to consider \emph{only} $\hat\Phi^{\ }_U$ because any global-$U(1)$ invariant observable can be put into alternating $\hat\psi_U^\dagger$, $\hat\psi^{\ }_V$ operator form, and, by polarization, using linearity and anti--linearity without using commutation relations,
\begin{eqnarray*}
  \hat\psi_U^\dagger\hat\psi^{\ }_V=\frac{1}{4}\left[\hat\Phi^{\ }_{U+V}-\hat\Phi^{\ }_{U-V}
                    -\rmi\hat\Phi^{\ }_{U+\rmi V}+\rmi\hat\Phi^{\ }_{U-\rmi V}\right],
\end{eqnarray*}
however it will often be convenient to use $\hat\psi_U^\dagger\hat\psi^{\ }_V$, particularly noting the projection property $(\hat\psi_U^\dagger\hat\psi^{\ }_V)^n=(U,V)^{n-1}\hat\psi_U^\dagger\hat\psi^{\ }_V$, more--or--less as for $\hat\Phi^{\ }_U$.
The properties of $\hat\Phi^{\ }_U$ ensure that it behaves rather like a decimation operator relative to lower--level observables $\hat\Phi_{U_i}$ (lower--level in the sense that $\mathrm{Supp}(U_i)\subset\mathrm{Supp}(U)$), with which it is in general incompatible.
The vacuum vector allows us to construct the vacuum state $\VEV{\cdot}$ over the algebra generated by $\hat\Phi^{\ }_U$ and non--vacuum states such as
\begin{eqnarray*}
\frac{\VEV{\hat\Phi^{\ }_{V_1}\cdots\hat\Phi^{\ }_{V_n}
                   \quad\cdot\quad\;\hat\Phi^{\ }_{V_n}\cdots\hat\Phi^{\ }_{V_1}}}
    {\VEV{\hat\Phi^{\ }_{V_1}\cdots\hat\Phi^{\ }_{V_n}\hat\Phi^{\ }_{V_n}\cdots\hat\Phi^{\ }_{V_1}}}.
\end{eqnarray*}
The intentional restriction to using only the observable operators $\hat\Phi^{\ }_{V}$ to generate states, which can be characterized as zero charge states, is not operationally significant, because we can always construct operators $\hat\psi^\dagger_U\hat\psi^{\ }_V$ for which either $U$ or $V$ is at arbitrarily large separation from the region of space--time that contains an experiment.

We can introduce a bosonic raising and lowering algebra that includes an observable that satisfies the same Lie algebra as is satisfied by $\hat\Phi^{\ }_U$,
\begin{eqnarray*}
  [\sfa^{\ }_U,\sfa^\dagger_V]=(U,V),\qquad [\sfa^{\ }_U,\sfa^{\ }_V]=0,
\end{eqnarray*}
for which $\hat\Chi^{\ }_U=\sfa^\dagger_{U^c}\sfa^{\ }_{U^c}$ satisfies
\begin{eqnarray*}
  [\hat\Chi^{\ }_U,\hat\Chi^{\ }_V]
       &=&[\sfa^\dagger_{U^c}\sfa^{\ }_{U^c},\sfa^\dagger_{V^c}\sfa^{\ }_{V^c}],\\
       &=&\sfa^\dagger_{U^c}[\sfa^{\ }_{U^c},\sfa^\dagger_{V^c}\sfa^{\ }_{V^c}]+
                  [\sfa^\dagger_{U^c},\sfa^\dagger_{V^c}\sfa^{\ }_{V^c}]\sfa^{\ }_{U^c}\\
       &=&(U^c,V^c)\sfa^\dagger_{U^c}\sfa^{\ }_{V^c}
                    - (V^c,U^c)\sfa^\dagger_{V^c}\sfa^{\ }_{U^c}\\
       &=&(V,U)\sfa^\dagger_{U^c}\sfa^{\ }_{V^c}
                    - (U,V)\sfa^\dagger_{V^c}\sfa^{\ }_{U^c}\\
    &=&\rmi\sqrt{(U,U)(V,V)}\Bigl[\hat\Chi^{\ }_{Y(V,U)}-\hat\Chi^{\ }_{Y(U,V)}\Bigl]
\end{eqnarray*}
allowing the identification $\hat\Chi^{\ }_U\equiv\hat\Phi^{\ }_U$, or, equivalently, $\sfa^\dagger_{U^c}\sfa^{\ }_{V^c}\equiv\hat\psi_U^\dagger\hat\psi^{\ }_V$
(with care taken to ensure equivalence of complex linearity and anti--linearity in $U$ and in $V$).
The construction so far can be compared with the Jordan-Wigner transformation\cite{tHooft}\hspace{-0.1em}(\S 15.1)
      ${}^{\hspace{-0.4em},}$\cite{JordanWigner}.
We can construct a real random field using $\sfa^{\ }_U$ and $\sfa^\dagger_U$,
\begin{eqnarray*}
  \hat\chi^{\ }_U=\sfa^{\ }_{U^c}+\sfa^\dagger_U,\qquad \hat\chi^{\dagger}_U=\hat\chi^{\ }_{U^c},
\end{eqnarray*}
with trivial commutator,
\begin{eqnarray*}
  [\hat\chi^{\ }_U,\hat\chi^{\ }_V]
         &=&[\sfa^{\ }_{U^c},\sfa^\dagger_V]+[\sfa^\dagger_U,\sfa^{\ }_{V^c}]\\
         &=&(U^c,V)-(V^c,U)=0.
\end{eqnarray*}
As a next step, we can extend the state we have over the algebra generated by the global-$U(1)$ invariant observables $\hat\Phi^{\ }_U$, which is therefore also a state over the $\hat\Chi^{\ }_U$, to be a state $\VEVx{\cdot}$ over the algebra generated by $\sfa^\dagger_{U^c}$ and $\sfa^{\ }_{V^c}$, which we will do here by the simplest possible prescription, that any term in an expanded expression that cannot be presented as a product of factors $\hat\Chi^{\ }_U$ will be assigned the value 0.
The notation $\VEVx{\cdot}$ intends to emphasize that the vacuum vector $\VEVxR{}$ is not annihilated by $\sfa^{\ }_{V^c}$, just as $\left|0\right>$ is not annihilated by $\hat\psi^{\ }_V$.

The resulting state can be fixed by constructing a generating function for the random field $\hat\chi^{\ }_U$,
\begin{eqnarray*}
\fl\mbox{(apply a Baker--Campbell--Hausdorff identity ...)}\\
\fl\hspace{2em}\VEVx{\rme^{\rmi\lambda\hat\chi^{\ }_U}}=
       \VEVx{\rme^{\rmi\lambda\sfa^{\dagger}_U}\rme^{\rmi\lambda\sfa^{\ }_{U^c}}}
                                     \rme^{-\lambda^2(U^c,U)/2}\\
\fl\mbox{(only include terms with equal numbers of 
                           $\sfa^\dagger_U$, $\sfa^{\ }_{U^c}$ ...)}\\
\fl\hspace{2em}=\VEVx{1+\sum\limits_{j=1}^\infty\frac{(-\lambda^2)^j}{j!^2}
                                      (\sfa^{\dagger}_U)^j(\sfa^{\ }_{U^c})^j}
                                     \rme^{-\lambda^2(U^c,U)/2}\\
\fl\mbox{(use the commutator $[\sfa^{\ }_{U^c},\sfa^\dagger_U]=(U^c,U)$, giving $\sfa^{\dagger j}_U\sfa^{\ }_{U^c}
                   {=}\Bigl(\sfa^{\dagger}_U\sfa^{\ }_{U^c}{-}(j{-}1)(U^c,U)\Bigr)\sfa^{\dagger (j-1)}_U$ ...)}\\
\fl\hspace{-2em}=\VEVxL{1+\sum\limits_{j=1}^\infty\frac{(-\lambda^2)^j}{j!^2}
                                      \Bigl(\sfa^{\dagger}_U\sfa^{\ }_{U^c}{-}(j{-}1)(U^c,U)\Bigr)}
\!\times\cdots\times\!
                   \VEVxR{\Bigl(\sfa^{\dagger}_U\sfa^{\ }_{U^c}{-}(U^c,U)\Bigr)\sfa^{\dagger}_U\sfa^{\ }_{U^c}}
                                     \rme^{-\lambda^2(U^c,U)/2}\\
\fl\mbox{(only at this point do we map $\sfa^{\dagger}_U\sfa^{\ }_{U^c}$ to $\hat\psi_{U^c}^\dagger\hat\psi^{\ }_{U}$ and $\VEVx{\cdot}$ to $\VEV{\cdot}$, ...)}\\
\fl\hspace{-2em}=\langle 0|1+\sum\limits_{j=1}^\infty\frac{(-\lambda^2)^j}{j!^2}
                                      \Bigl(\hat\psi_{U^c}^\dagger\hat\psi^{\ }_{U}{-}(j{-}1)(U^c,U)\Bigr)
\!\times\,\cdots\,\times\!
        \Bigl(\hat\psi_{U^c}^\dagger\hat\psi^{\ }_{U}{-}(U^c,U)\Bigr)
                             \hat\psi_{U^c}^\dagger\hat\psi^{\ }_{U}|0\rangle
                                     \rme^{-\lambda^2(U^c,U)/2}\\
\fl\mbox{(only the constant and the $-\lambda^2$ terms survive ...)}\\
\fl\hspace{2em}=(1-\lambda^2(U^c,U)_+)\rme^{-\lambda^2(U^c,U)/2}
\end{eqnarray*}
{\small$\big[$more generally, we have $\VEVx{\rme^{\rmi\lambda\sfa^{\ }_{U'}+\rmi\mu\sfa^{\dagger}_U}}
             =(1-\lambda\mu(U',U)_+)\rme^{-\lambda\mu(U',U)/2}$ as a generating function for the whole raising and lowering algebra$\big]$.}
This should be compared with the generating function for the conventional Gaussian vacuum state $\VEVchi{\cdot}$,
which for $\hat\chi^{\ }_U$ would be
\begin{eqnarray*}
  \VEVchi{\rme^{\rmi\lambda\hat\chi^{\ }_U}}=\rme^{-\lambda^2(U^c,U)/2}.
\end{eqnarray*}
For any test function $V$, we can construct a first degree raised state
\begin{eqnarray*}
  \frac{\VEVchi{\hat\chi^\dagger_V\rme^{\rmi\lambda\hat\chi^{\ }_U}\hat\chi^{\ }_V}}
        {\VEVchi{\hat\chi^\dagger_V\hat\chi^{\ }_V}}
             =\left[1-\lambda^2\frac{(U^c,V)(V,U)}{(V,V)}\right]\rme^{-\lambda^2(U^c,U)/2},
\end{eqnarray*}
so we can consider $\VEVx{\cdot}$ to be an equally weighted convex mixture of this state for all test functions $V$ for which $(V,V)_+=(V,V)=1$, so that $\VEVx{\cdot}$ is unitarily inequivalent to $\VEVchi{\cdot}$.
When $U^c=U$, $\hat\chi^{\ }_U$ is Hermitian, so we obtain a probability density for single measurements, by inverse fourier transform,
\begin{eqnarray*}
  \VEVx{\delta(\hat\chi^{\ }_U - v)}=
                  \left[\frac{(U,U)_-}{(U,U)}+\frac{(U,U)_+}{(U,U)}v^2\right]
                         \frac{\rme^{\textstyle-\frac{v^2}{2(U,U)}}}{\sqrt{2\pi(U,U)}},
\end{eqnarray*}
varying continuously between a Gaussian probability density and the globally raised second degree probability density, depending on the ratio $(U,U)_+/(U,U)_-$.

Despite this continuous probability density, for $\hat\Chi^{\ }_U=\hat\Chi^{\dagger}_U$ we obtain a generating function
\begin{eqnarray*}
  \VEVx{\rme^{\rmi\lambda\hat\Chi^{\ }_U}}&=&\VEV{\rme^{\rmi\lambda\hat\Phi^{\ }_U}}
    =\VEV{\left(1-\frac{\hat\Phi^{\ }_U}{(U,U)}
                                +\rme^{\rmi\lambda(U,U)}\frac{\hat\Phi^{\ }_U}{(U,U)}\right)}\\
    &=&\frac{(U,U)_-}{(U,U)}+\frac{(U,U)_+}{(U,U)}\rme^{\rmi\lambda(U,U)},
\end{eqnarray*}
from which we obtain a discrete two--valued probability density, isolated at $0$ and at $(U,U)$,
\begin{eqnarray*}
  &&\hspace{-1.5em}\VEVx{\delta(\hat\Chi^{\ }_U {-} v)}{=}
                  \frac{(U,U)_-}{(U,U)}\delta(v)+\frac{(U,U)_+}{(U,U)}\delta(v{-}(U,U)).
\end{eqnarray*}
This probability density for $\hat\Chi^{\ }_U\equiv\hat\Phi^{\ }_U$ in the ground state of the fermionic field is conventional enough that by the usual rules of quantum mechanics we ought to be able to measure it, however the probability density for the Dirac spinor random field $\hat\chi_U$ is somewhat an extrapolation.
We can consistently suppose or imagine that there might be such a random field as $\hat\chi_U$, and a state $\VEVx{\cdot}$ over those observables, behind the scenes of the observables $\hat\Chi^{\ }_U$, even if we can never measure them, or even if we never try to measure them, but we could also think of them as a challenge, that such a field might be there to be measured if we can think of a way to do it.
We can also take $\VEVx{\cdot}$ as an inspiration for other bosonic fields, so that we could, for example, introduce a similar Poincar\'e invariant state for the quantized electromagnetic field, using the ratio $(\sff,\Half(1{-}\rmi\star)\sff)_+/(\sff,\Half(1{+}\rmi\star)\sff)_+$ instead of the ratio $(U,U)_+/(U,U)_-$,
$$\VEV{\delta(\hat\sfF_{\sff} - v)}=
                  \left[\frac{(\sff,\Half(1{-}\rmi\star)\sff)_+}{(\sff,\sff)_+}+\frac{(\sff,\Half(1{+}\rmi\star)\sff)_+}{(\sff,\sff)_+}v^2\right]
                         \frac{\rme^{\textstyle-\frac{v^2}{2(\sff,\sff)_+}}}{\sqrt{2\pi(\sff,\sff)_+}},
$$
where, as usual, we require $\sff^*\,{=}\,\sff$ so that $\hat\sfF_{\sff}^\dagger\,{=}\,\hat\sfF_{\sff}$ is observable.
This state for the quantized electromagnetic field is too exaggeratedly chiral to be physical without being considerably mollified as a mixture with the usual electromagnetic vacuum state, but such a construction would likely not be thought of without this kind of investigation \emph{as} an inspiration.

For any state generated by the action of $\hat\Phi^{\ }_V$ on $\VEV{\cdot}$ we have equivalences between generating functions such as
$$\frac{\VEV{\hat\Phi^{\ }_{V_1}\cdots\hat\Phi^{\ }_{V_n}
                   \rme^{\rmi\lambda\hat\Phi^{\ }_U}\hat\Phi^{\ }_{V_n}\cdots\hat\Phi^{\ }_{V_1}}}
    {\VEV{\hat\Phi^{\ }_{V_1}\cdots\hat\Phi^{\ }_{V_n}\hat\Phi^{\ }_{V_n}\cdots\hat\Phi^{\ }_{V_1}}}
=  \frac{\VEVx{\hat\Chi^{\ }_{V_1}\cdots\hat\Chi^{\ }_{V_n}
                   \rme^{\rmi\lambda\hat\Chi^{\ }_U}\hat\Chi^{\ }_{V_n}\cdots\hat\Chi^{\ }_{V_1}}}
    {\VEVx{\hat\Chi^{\ }_{V_1}\cdots\hat\Chi^{\ }_{V_n}\hat\Chi^{\ }_{V_n}\cdots\hat\Chi^{\ }_{V_1}}},
$$
the latter of which can be extended to a generating function for $\hat\chi^{\ }_U$,
$$\hspace*{-2em}
\frac{\VEVx{\hat\Chi^{\ }_{V_1}\cdots\hat\Chi^{\ }_{V_n}
                   \rme^{\rmi\lambda\hat\chi^{\ }_U}\hat\Chi^{\ }_{V_n}\cdots\hat\Chi^{\ }_{V_1}}}
    {\VEVx{\hat\Chi^{\ }_{V_1}\cdots\hat\Chi^{\ }_{V_n}\hat\Chi^{\ }_{V_n}\cdots\hat\Chi^{\ }_{V_1}}}
=\left[1-\lambda^2\frac{\VEV{\hat\Phi^{\ }_{V_1}\cdots\hat\Phi^{\ }_{V_n}
                                              \hat\psi^{\dagger}_{U^c}\hat\psi^{\ }_{U}
                                              \hat\Phi^{\ }_{V_n}\cdots\hat\Phi^{\ }_{V_1}}}
                                    {\VEV{\hat\Phi^{\ }_{V_1}\cdots\hat\Phi^{\ }_{V_n}
                                              \hat\Phi^{\ }_{V_n}\cdots\hat\Phi^{\ }_{V_1}}}\right]
             \rme^{-\lambda^2(U^c,U)/2}.
$$
For a state generated by the action of $\hat\Phi^{\ }_V$, for example, we obtain a generating function
\begin{eqnarray*}
\fl\frac{\VEVx{\hat\Chi^{\ }_{V}\rme^{\rmi\lambda\hat\chi^{\ }_U}\hat\Chi^{\ }_{V}}}
                                {\VEVx{\hat\Chi^{\ }_{V}\hat\Chi^{\ }_{V}}}
     &=&\left[1-\lambda^2\frac{\VEV{\hat\Phi^{\ }_{V}
                                              \hat\psi^{\dagger}_{U^c}\hat\psi^{\ }_{U}
                                              \hat\Phi^{\ }_{V}}}
                                    {\VEV{\hat\Phi^{\ }_{V}\hat\Phi^{\ }_{V}}}\right]
             \rme^{-\lambda^2(U^c,U)/2}\\
\fl     &=&\left[1{-}\lambda^2\!\left(\!(U^c{\perp}V,U{\perp}V)_+  +  \frac{(U^c,V)(V,U)}{(V,V)}
                                         \!\right)\right]\rme^{-\lambda^2(U^c,U)/2},
\end{eqnarray*}
where $U{\perp}V=U-\frac{(V,U)}{(V,V)}V$ (see \ref{CalculateVEVs} for this calculation in more detail); for a state generated by the action of $\hat\psi^{\dagger}_V\hat\psi^{\ }_W$ (with $V$ and $W$ orthogonal, $(V,W)=0$, to simplify the expression, which in particular will be satisfied whenever $V$ and $W$ have space--like separated supports),
\begin{eqnarray*}
\fl\frac{\VEVx{\sfa^{\dagger}_{W^c}\sfa^{\ }_{V^c}\rme^{\rmi\lambda\hat\chi^{\ }_U}
                                                                \sfa^{\dagger}_{V^c}\sfa^{\ }_{W^c}}}
                                {\VEVx{\sfa^{\dagger}_{W^c}\sfa^{\ }_{V^c}\sfa^{\dagger}_{V^c}\sfa^{\ }_{W^c}}}
     =\left[1-\lambda^2\frac{\VEV{\hat\psi^{\dagger}_{W}\hat\psi^{\ }_{V}
                                              \hat\psi^{\dagger}_{U^c}\hat\psi^{\ }_{U}
                                              \hat\psi^{\dagger}_{V}\hat\psi^{\ }_{W}}}
                                    {\VEV{\hat\psi^{\dagger}_{W}\hat\psi^{\ }_{V}
                                              \hat\psi^{\dagger}_{V}\hat\psi^{\ }_{W}}}\right]
             \rme^{-\lambda^2(U^c,U)/2}\\
\fl\qquad   =\left[1-\lambda^2\left(\!\bigl(U^c{\perp}V{\perp}W,U{\perp}V{\perp}W\bigr)_+
+  \frac{(U^c,V)(V,U)}{(V,V)}\!\right)\right]\rme^{-\lambda^2(U^c,U)/2}\\
\noalign{\centerline{\vspace{-5.5ex}[where we can omit brackets when $(V,W)=0$, so that $U{\perp}V{\perp}W=(U{\perp}V){\perp}W=(U{\perp}W){\perp}V$];}}\\
\end{eqnarray*}
and for a state generated by the action of $\hat\Phi^{\ }_V\hat\Phi^{\ }_W$ (again with $V$ and $W$ orthogonal),
\begin{eqnarray*}
\fl\frac{\VEVx{\hat\Chi^{\ }_{W}\hat\Chi^{\ }_{V}\rme^{\rmi\lambda\hat\chi^{\ }_U}
                                               \hat\Chi^{\ }_{V}\hat\Chi^{\ }_{W}}}
         {\VEVx{\hat\Chi^{\ }_{W}\hat\Chi^{\ }_{V}\hat\Chi^{\ }_{V}\hat\Chi^{\ }_{W}}}\cr
\fl\quad=\left[1-\lambda^2\left(\!\left(U^c{\perp}V{\perp}W,U{\perp}V{\perp}W\right)_+
        +  \frac{(U^c,V)(V,U)}{(V,V)}          +  \frac{(U^c,W)(W,U)}{(W,W)}
                                         \!\right)\right]\rme^{-\lambda^2(U^c,U)/2};
\end{eqnarray*}
so that $\hat\Phi^{\ }_V$, $\hat\psi^{\dagger}_V\hat\psi^{\ }_W$, $\hat\Phi^{\ }_{V}\hat\Phi^{\ }_{W}$, and higher degree operators act to modulate the vacuum state in a limited way, always with a factor $1{-}\lambda^2\mathsf{M}(U^c,U)$, linear in $U^c$ and in $U$, which can be reproduced by an appropriately chosen mixture of actions of $\hat\chi^{\ }_V$ on the vacuum state $\VEVchi{\cdot}$; we could equally well say that such states are generated by a constrained action of the algebra generated by $\hat\chi^{\ }_V$ and call them classical.

The fermionic Hilbert space does not contain all the states we can construct using $\hat\chi^{\ }_U$ acting on $\VEVxR{}$ or on $\VEVchiR{}$ (just as superselection prohibits many vectors that we could construct using the unconstrained action of $\hat\psi^{\ }_U$ and $\hat\psi^{\dagger}_U$ on $|0\rangle$, allowing superpositions only of vectors of equal global-$U(1)$ charge), instead being restricted only to the states we can construct using many different $\hat\Chi^{\ }_{V}$ acting on $\VEVxR{}$.
In effect, we are only able to construct states ---\,from the vacuum state we are given as a starting point--- using the projective quantum mechanical measurements we can actually perform, such as the $\hat\Chi^{\ }_U\equiv\hat\Phi^{\ }_U$, suitably normalized, so that $\hat\Chi^{\ 2}_U=\hat\Chi^{\ }_U$ or equivalently $\hat\Phi^{\ 2}_U=\hat\Phi^{\ }_U$, corresponding to the yes/no sample space of waiting for a detector somewhere to click.
We \emph{can} construct all the necessary states using the random field $\hat\chi^{\ }_U$, but the constraint is perhaps not at this point classically well--motivated enough for us to say that such states are truly ``classical''.

\section{$U(1)$--gauge invariant observables\label{LocalU1}}
\newcommand\pmm{{\!+\!\!\diagup\!\!\!-}}
We here only indicate a possible approach to $U(1)$--gauge invariance.
Under $U(1)$--gauge transformations, the Dirac spinor wave function transforms as
$\hat\psi_\xi(x)\mapsto\rme^{\rmi\theta(x)}\hat\psi_\xi(x)$\cite[Eq. 2-63]{IZ}, so that the vacuum expectation values
$\iS_{+\xi\xi'}(x,x')=\VEV{\hat\psi_\xi(x)\overline{\hat\psi_{\xi'}(x')}}$ and
$\iS_{-\xi\xi'}(x,x')=\VEV{\overline{\hat\psi_{\xi'}(x')}\hat\psi_\xi(x)}$
transform as parallel transports,
$$\iS_{\pm\xi\xi'}(x,x')\mapsto\rme^{\rmi(\theta(x)-\theta(x'))}\iS_{\pm\xi\xi'}(x,x').$$
We can therefore construct two--point $U(1)$--gauge invariant operators (for both free and interacting fields, supposing interacting fields exist), using these parallel transports and two Dirac matrix--valued test functions, $P_{\xi'\eta'}(x')$ and $Q_{\eta\xi}(x)$,
$$\sum\limits_{\xi'\eta'\eta\xi}\int\overline{\hat\psi_{\xi'}(x')}P_{\xi'\eta'}(x')\iS_{\pm\eta'\eta}(x',x)Q_{\eta\xi}(x)\hat\psi_{\xi}(x)
  \Intd^4x\Intd^4x',$$
which we will write without indices as
$\int\overline{\hat\psi(x')}P(x')\iS_{\pm}(x',x)Q(x)\hat\psi(x)\Intd^4x\Intd^4x'$.
Of the possible linear combinations of these two constructions,
$\int\overline{\hat\psi(x')}P(x')\iS_{\pmm}(x',x)Q(x)\hat\psi(x)\Intd^4x\Intd^4x'$, where $\iS_{\pmm}(x',x)=\iS_{+}(x',x)-\iS_{-}(x',x)$, is notably less singular on the light--cone in the free field case, indeed this is the least singular $U(1)$--gauge invariant construction known to the author; in particular, it is less singular than constructions that use the electromagnetic potential operator.
The appearance of Dirac matrix--valued test functions $P(x')$ and $Q(x)$ in this $U(1)$--gauge invariant construction lessens the significance of the double cover of the Lorentz group.
The construction $\int\overline{\hat\psi(x')}P(x')\iS_{\pmm}(x',x)Q(x)\hat\psi(x)\Intd^4x\Intd^4x'$ cannot be written as a simple product $\hat\psi_U^\dagger\hat\psi_U^{\,}$, nonetheless it can be written in terms of bosonic raising and lowering operators.

\section{Discussion}
The constructions above, for free quantum fields, which only apply where interactions are taken to be insignificant ---that is, to the in-- and out--states of the S--matrix and to simple quantum optics--- do not touch on how we might discuss interactions in classical canonical terms as well as or instead of in quantum unitary terms.
The introduction of negative frequency components by the constructions here of random fields makes it impossible to preserve the Correspondence Principle, which includes correspondence between 4--momentum and wave--number, $p=\hbar k$, with the energy component $p_0$ required to be positive semi--definite.
The Correspondence Principle, however, can be regarded as a property of the quantization process, not of the Hilbert space that is created by quantization of a classical dynamics, whereas the Koopman--von Neumann approach described in \ref{CMtoQM} is a different process for constructing the same Hilbert space from a different classical dynamics, for which frequency is not related in the same way to energy.

As noted in \cite{MorganEMunpublished}, there has been discussion of the similarities and differences between ``random electrodynamics'' and quantum electrodynamics at least since the 1960s\cite{BoyerA,BoyerB}, however the algebraic formalism used here makes the comparison and the establishment of empirical equivalence much more direct.
The negative frequencies that appear explicitly in the algebraic approach here appear implicitly in random electrodynamics (which has come to be called ``Stochastic Electrodynamics'' or ``SED''\cite{SED}) as a factor $\cos{\hspace{-0.1em}(\mathbf{k}{\cdot}\mathbf{r}-\omega t)}$ in the 2--point correlation functions of the random electromagnetic field\cite[Eqs. (10), (13), and (14)]{BoyerA}, instead of a factor $\rme^{\rmi(\mathbf{k}{\cdot}\mathbf{r}-\omega t)}$ in the 2--point vacuum expectation values of the quantized electromagnetic field\cite[Eqs. (15), (16), and (17)]{BoyerA}.
The positive spectrum condition in quantum field theory can be thought of loosely as an \emph{a priori} requirement for a complex analytic structure, in contrast to an alternative requirement, for better or worse for different uses, for a classically real structure.

Although the main text has used a manifestly Lorentz invariant 4--dimensional block world formalism, if we choose a time--like 4--vector we can construct a 3--dimensional formalism by reducing
$$2\pi\delta(k{\cdot}k-m^2)\frac{\Intd^4k}{(2\pi)^4}\quad\mbox{to}\quad 
2\pi\frac{\delta(k_0-\sqrt{\mathbf{k}{\cdot}\mathbf{k}+m^2})
        +\delta(k_0+\sqrt{\mathbf{k}{\cdot}\mathbf{k}+m^2})}
           {2\sqrt{\mathbf{k}{\cdot}\mathbf{k}+m^2}}\frac{\Intd^4k}{(2\pi)^4},$$
and hence reduce the 4--dimensional formalism to a 3--dimensional formalism, so we do not have to interpret the formalism as requiring a 4--dimensional block world ontology.

Given quantum states, the properties of which are determined as far as possible by using quantum field measurements, it is of course possible to ask what the results of classical measurements would be, as indeed we did above for the Dirac spinor--valued random field, if only we could make those classical measurements or if we could correct for whatever a classical theory takes to be, from its perspective, the inadequacies of the quantum measurements\cite{TsangCaves}.
From a classical perspective, we can take a presentation of what the results of such classical measurements would be to constitute a different way to work with the Poincar\'e invariant noise of the vacuum state.
To briefly address only two of very many contemporary issues:
\begin{itemize}
\item a classical perspective takes quantum computation and other exploitation of the quantum-Hilbert space formalism to be a consequence of the measurement process, the reduction of elaborate noisy classical states by the use of local and nonlocal observables that have discrete spectra and incompatible eigenspaces, however this is only a matter of interpretation, because the mathematical landscape provided by field operators acting on the vacuum sector Hilbert space is completely unchanged;
\item the random fields of quantum non-demolition operators constructed here make it possible to discuss the violation of Bell inequalities by random fields and by quantum fields in a unified way, which is somewhat anticipated in \cite[\S 6]{MorganBellRF}, insofar as any quantum field state is equivalent to a nonlocally correlated, noisy, but nonetheless effectively superdetermined classical random field state, however antilocality of the propagators involved\cite{SegalGoodman,Hegerfeldt}, microcausality, and noncommutativity of the algebra of observables\cite{Landau} must also be considered.
\end{itemize}

Particularly for the quantized Dirac spinor field, however, the projective quality of many measurements limits how closely we might determine a putative underlying classical state, so that we may well be best to discuss quantum measurements as about quantum states, with only a background acknowledgment that perhaps there are classical random fields underpinning all this, or perhaps there are not, with no prejudicial determination either way.
From a practical point of view, the Correspondence Principle is so embedded in physicists' thinking that it will be best to keep thinking in terms of both quantum fields and random fields.

Finally, the focus on the space of test functions and its pre--inner product structure is very much aimed towards future consideration of interacting quantum or random fields in test function algebraic terms, insofar as products and derivatives of test functions are always well--defined, steps towards which may be found in \cite{MorganTF}, in contrast to using renormalization to fix the problems introduced by defining a Lagrangian evolution using products of distributions.
For free quantum fields, expected values for quantum field operators depend linearly on the modulation that is applied to the vacuum.
From a signal analysis perspective, that is a convenience more than a necessity, surely not definite enough for it to be enshrined in axioms (as it is directly by the Wightman axioms, by requiring quantum fields to be distributions, but only indirectly by the Haag-Kastler axioms, through additivity).
Response to modulations is in general not linear in physics (except as a first approximation or because we engineer the response to be linear over as large a range as we can.)
Thus, we might usefully introduce nonlinear dependence on test functions (which is well-defined) instead of introducing powers of distributions as interaction terms (which is not).
If we take the renormalization scale that is required to construct interacting theories to be fixed by or at least to be correlated with parameters of the test functions used in detailed models of an experiment, then interacting theories are already weakly nonlinearly dependent on the test functions.

Better understanding the relationship between random and quantum fields, by isomorphisms as well as by quantization, allows us more to apply a classical intuition to the construction of interacting field theories.
For example, when thinking classically an introduction of a Dirac spinor field could be understood to be a way to model dispersion, nonlinearity, and constraints of the electromagnetic field, not necessarily as a new particle, or else we could eliminate the electromagnetic field; ideas are more fluid in the classical setting.
Even though we can't work classically with complete freedom, because there are isomorphisms only for \emph{some} quantum field models, nonetheless the limits of when classical intuition might be useful are well-defined by what isomorphisms we can construct.

\ack
I am grateful for conversation with Carlton Caves and for correspondence with Federico Zalamea, Jess Riedel, Jean-Pierre Magnot, Joseph Eberly, David Alan Edwards, Stephen Paul King, and Michael Hall.

\appendix
\section{Classical mechanics to quantum mechanics: Koopman--von Neumann\label{CMtoQM}}
We can present Classical Mechanics as a commutative, associative algebra $\mathcal{A}$ of observables over a phase space, $A:\mathcal{P}\rightarrow\mathbb{R}; P\mapsto A(P)$, with a multiplication $\cdot{:}\mathcal{A}{\times}\mathcal{A}{\rightarrow}\mathcal{A};A,B\mapsto A\cdot B$, \emph{together with} a Poisson bracket, $\{\bullet,\bullet\}{:}\mathcal{A}{\times}\mathcal{A}{\rightarrow}\mathcal{A}:A,B\mapsto\{A,B\}$, and a Hamiltonian $H(P)$ that generates, for an observable $A(P,t)$, an evolution $\frac{dA}{dt}=\{H,A\}+\frac{\partial A}{\partial t}$.
[{\small We can present Classical Mechanics equivalently, with considerably more elaborate machinery, using the Peierls bracket over the solution space instead of using the Poisson bracket over phase space\cite[\S 4.4.1]{Rejzner}, however this \appendixname~will use the simpler machinery of phase space.}]
The algebra of classical observables is \emph{not} a simple commutative algebra.

We can use the multiplication $\cdot$ to construct an action
$$\hat Y_A:\mathcal{A}\rightarrow\mathcal{A};\bullet\mapsto \hat Y_A(\bullet)=A\cdot\bullet,$$
where we can identify the algebra generated by the $\hat Y_A$ with $\mathcal{A}$, and we can similarly use the Poisson bracket to construct what can be called \emph{generators of transformations},
$$\hat Z_A:\mathcal{A}\rightarrow\mathcal{A};\bullet\mapsto \hat Z_A(\bullet)=\{A,\bullet\},$$ which act noncommutatively but associatively on $\mathcal{A}\,$\cite{Koopman,vonNeumann},\cite{tHooft}\hspace{-0.1em}(\S\S 2.1.1,\,5.5.1),\cite{Woodhouse}\hspace{-0.1em}(\S\S 1.5-6),\cite{Zalamea,Kisil,Mauro,deGossonA,deGossonB,Blasone,Klein,Rajagopal,Viennot}.
These two actions allow us to construct a noncommutative, associative algebra $\mathcal{A}_+$ that is generated by the $\hat Y_A$ \emph{and} the $\hat Z_A$, satisfying the commutation relations $[\hat Y_A,\hat Y_B]=0$, $[\hat Z_A,\hat Y_B]=\hat Y_{\{A,B\}}$, and $[\hat Z_A,\hat Z_B]=\hat Z_{\{A,B\}}$.
The linear space of operators generated by the $\hat Y_A$ supports an adjoint action of the Lie algebra generated by the $\hat Z_A$.
Note that this algebra of operators has always been implicitly part of a Hamiltonian presentation of classical mechanics, even though it has been explicitly presented by the Poisson bracket.
If the phase space is elementary, for a dynamics that is either without constraints or for which all constraints are implemented by the Hamiltonian, $\mathcal{A}_+$ is generated by $q_i$, $p_i$, $\partial/\partial q_i$, and $\partial/\partial p_i$.\vspace{1ex}

The above is all just Classical Mechanics. To move towards Quantum Mechanics, we introduce raising and lowering operators for the elementary case,
\begin{eqnarray*}
  a_i^\dagger=\frac{1}{\sqrt{2}}\left(q_i-\frac{\partial}{\partial q_i}\right),&\qquad&
  b_i^\dagger=\frac{1}{\sqrt{2}}\left(p_i-\frac{\partial}{\partial p_i}\right),\\
  a_i=\frac{1}{\sqrt{2}}\left(q_i+\frac{\partial}{\partial q_i}\right),&\qquad&
  b_i=\frac{1}{\sqrt{2}}\left(p_i+\frac{\partial}{\partial p_i}\right),
\end{eqnarray*}
which, when taken with $(\hat W\hat X)^\dagger{=}\hat X^\dagger \hat W^\dagger$, define an involution $\hat X{\mapsto}\hat X^\dagger$, making $\mathcal{A}_+$ a $*$--algebra over $\mathbb{R}$ (that is, we allow only real scalar multiples).
The raising and lowering operators satisfy the commutation relations $[a_i,a_j^\dagger]=\delta_{ij}$, $[b_i,b_j^\dagger]=\delta_{ij}$, and $[a_i,a_j]{=}[a_i,b_j]{=}[b_i,b_j]{=}[a_i,b_j^\dagger]{=}0$.
If we set $\rho(1)=1$ and
$$\rho(\hat Xa_i)=\rho(\hat Xb_i)=\rho(a_i^\dagger \hat X)=\rho(b_i^\dagger \hat X)=0\quad\forall \hat X\in\mathcal{A}_+,$$
then we obtain by the usual manipulations a Gaussian statistical state\cite[\S III.2.2]{Haag} over $\mathcal{A}_+$, $\rho:\mathcal{A}_+\rightarrow\mathbb{C}$, which is a complex linear form that satisfies $\rho(\hat X^\dagger \hat X)\ge 0$, $\rho(1)=1$, and $\rho(\hat X^\dagger)=\overline{\rho(\hat X)}$, which allows a probability interpretation for those operators for which $\hat X^\dagger=\hat X$ (which we can therefore call ``observables''), and which allows the GNS-construction\cite[\S III.2.2]{Haag} of a Hilbert space $\mathcal{H}_+$.

The presence or otherwise of a natural complex structure makes a difference\cite{Kisil}.
$\mathcal{H}_+$ as a vector space can be generated by a basis that contains a vacuum vector $|0\rangle$ and \emph{real}--valued multiples of $a_1^{\dagger j_1}b_1^{\dagger k_1}\cdots a_m^{\dagger j_m}b_m^{\dagger k_m}|0\rangle$.
Classical mechanical systems that have a natural complex structure ---which for a random field can be provided, as in Section \ref{EMfield}, by the cosine and sine components of the fourier transform relative to space--time coordinates--- are equivalent to a system of quantized simple harmonic oscillators, which can be generated by a basis that contains a vacuum vector $|0\rangle$ and \emph{complex}--valued multiples of $a_1^{\dagger j_1}b_1^{\dagger k_1}\cdots a_m^{\dagger j_m}b_m^{\dagger k_m}|0\rangle$.
If we introduce an engineering imaginary $\Ej$ for the purposes of signal processing (though it is not necessary if we restrict ourselves to using fourier sine and cosine transforms) then each $q_i$,$\,\Ej\partial/\partial q_i$ and $p_i$,$\,\Ej\partial/\partial p_i$ pair allows the construction of a Wigner or other quasi--probability distribution, using the mathematics and classical measurement theory associated with time--frequency distributions\cite{Cohen} as $q_i$--$q_i\!$frequency distributions.

The positive definite Hamiltonian function for a collection of non-interacting simple harmonic oscillators, in a vector notation, with $\underline{q}=(q_1, q_2, ...)$ and $\underline{p}=(p_1, p_2, ...)$, is
$H(\underline{q},\underline{p})=\Half(\underline{q}\cdot\underline{q}+\underline{p}\cdot\underline{p})$, from which we obtain two operators, using $\underline{a}$ and $\underline{b}$ as vectors of raising and lowering operators, as we did for $\underline{q}$ and $\underline{p}$,
$$\hat Y_H={\scriptstyle\frac{\scriptstyle 1}{\scriptstyle 4}}\left[
      \bigl(\underline{a}{+}\underline{a}^\dagger\bigr){\cdot}\bigl(\underline{a}{+}\underline{a}^\dagger\bigr)
    +\bigl(\underline{b}{+}\underline{b}^\dagger\bigr){\cdot}\bigl(\underline{b}{+}\underline{b}^\dagger\bigr)
      \right],\qquad
  \hat Z_H=\underline{a}\cdot\underline{b}^\dagger-\underline{a}^\dagger\cdot\underline{b}.
$$
which are Hermitian and anti--Hermitian respectively.
Classical physics requires only that $\hat Y_H$ is bounded below, not that $\hat Z_H$ is positive.
$\hat Z_H$ generates time--like translations; given a complex structure $\Ej$, we can transform to the basis $\underline{c}=(\underline{a}+\Ej\underline{b})/\sqrt{2}$,
$\underline{d}=(\underline{a}-\Ej\underline{b})/\sqrt{2}$, to obtain $\hat Z_H=\Ej\hat H_c$, where
$\hat H_c=\underline{c}^\dagger\cdot\underline{c}-\underline{d}^\dagger\cdot\underline{d}$, so that from the perspective of quantum mechanics the Hamiltonian operator $\hat H_c$ is not positive--definite (however we see in the main text that we can transform negative frequency components into positive frequency components for at least some random field constructions; for an example of the appearance of this operator as a generator of time--like translations in quantum optics, see \cite[Eq. (12)]{TsangCaves}).
The construction of the involution and state above is notable, however, for fixing statistics directly instead of assuming or requiring that a Hamiltonian or a stochastic dynamics is available to define, for example, a Gibbs state.
We have become accustomed to presenting classical physics using a Lagrangian or Hamiltonian, however we can in a stochastic context present classical physics using a state over an abstract $*$--algebra of observables.

The whole process here is an algebraic form of a Koopman--von Neumann approach, having four steps:
(1) use the injection $\mathcal{P}:\mathcal{A}\hookrightarrow\mathcal{A}_+$;
(2) introduce an involution $\hat X{\mapsto}\hat X^\dagger$, making $\mathcal{A}_+$ a $*$--algebra;
(3) introduce a statistical state over $\mathcal{A}_+$;
(4) use the GNS-construction of the Hilbert space $\mathcal{H}_+$.
Only (2) and (3) need the introduction of new structure, an involution and a state, for either of which there may be obstructions for more elaborate phase spaces; (1) and (4) use structure that's already there.
It is clear that this Koopman--von Neumann approach is quite different from canonical quantization, a map that is not an algebra morphism that for elementary cases can be presented as
$\mathcal{Q}:\mathcal{A}\rightarrow\mathcal{A}'\subset\mathcal{A}_+;(q_i,p_i)\mapsto(q_i,-\rmi\partial/\partial q_i)$.
At the level of quantum field theory, this should cause little concern, insofar as the choice of a classical field theory to quantize is rather instrumental: we choose a classical field to quantize that gives by quantization a quantum field that is empirically successful.
As an instrumental process, if there is an empirically successful quantum field that is the result of a Koopman--von Neumann treatment of some classical field, we can choose to use that classical field;  \cite{MorganEPL} (and \ref{complexQKG}) and Section \ref{EMfield} effectively show that there is such a classical field for the quantized complex Klein-Gordon field and for the quantized electromagnetic field, respectively.

\section{The complex Klein-Gordon field}\label{complexQKG}
To emphasize the similarity between the quantized electromagnetic field and the complex Klein-Gordon quantum field, instead of relying on the nonunique and rather dissimilar construction in \cite{MorganEPL}, we can present the complex Klein-Gordon quantum field as
$\hat\sfF_{\!f}=a^{\ }_{f^*}+a_f^\dagger=\hat\phi^{\ }_{f_1}+\hat\phi^\dagger_{f_2^*}$,
where $f=\left(f_1\atop f_2 \right)$ is a two component test function,
$f^*=\left(f_2^*\atop f_1^* \right)$ ensures that $\hat\sfF_{\!f}^\dagger=\hat\sfF^{\ }_{\!f^*}$,
and the raising and lowering operators satisfy $[a_f,a_g^\dagger]=(f,g)_+$, with the pre--inner product
$$(f,g)_\pm=\int\!\left(\widetilde{f_1}^{\!*}\hspace{-0.2em}(k)\widetilde{g_1}(k)
                             +\widetilde{f_2}^{\!*}\hspace{-0.2em}(k)\widetilde{g_2}(k)\right)
        2\pi\delta(k{\cdot}k-m^2)\theta(\pm k_0)\frac{\Intd^4k}{(2\pi)^4}.$$
For this quantum field, we can use the matrix $I=\left(\!{\ \;0\ \;1\atop -1\ 0}\right)$ in the same way as the Hodge dual was used for the quantized electromagnetic field in Section \ref{EMfield} to construct an involution
${}^\bullet:\mathcal{S}\rightarrow\mathcal{S};
f\mapsto f^\bullet=\Half(1+\rmi I)f+\Half(1-\rmi I)f^-$, $f^{\bullet\bullet}=f$, so that (because as for the quantized electromagnetic field we have the identities $(f^*,g^*)_\pm=(g,f)_\mp$ and $(f^-,g^-)_\pm=(f,g)_\mp$) we can derive $(f^{*\bullet},g^\bullet)_\pm=(g^{*\bullet},f^\bullet)_\pm$, and hence we can construct a random field $\bbFF_{\!f}=a^{\ }_{f^{*\bullet}}+a^\dagger_{f^\bullet}$ for which $[\bbFF_{\!f},\bbFF_g]=0$.

\section{The electromagnetic potential\label{EMpotential}}
\newcommand{\Aip}[2]{{(\!(#1,#2)\!)}}
The electromagnetic potential does not ``play~nice'' with the helicity projection used in the main text.

For the quantized electromagnetic field, we have
$\VEV{\hat\sfF^\dagger_{\sff}\hat\sfF_{\sfg}}=(\sff,\sfg)_+=\Aip{\delta\sff}{\delta\sfg}_+$, where for 1--forms $u$ and $v$ we have the sesquilinear forms
$$\Aip{u}{v}_\pm=   -\hbar\!\int\!\tilde{u}^*(k){\cdot}\tilde{v}(k)
                         2\pi\delta(k{\cdot}k)\theta(\pm k_0)\frac{\Intd^4k}{(2\pi)^4},$$
which are not positive semi--definite in general but are positive semi--definite for $\Aip{\delta\sff}{\delta\sfg}_\pm$.
We have, therefore, for arbitrary 3--forms $u_{(3)}$ and 1--forms $u_{(1)}$,
\begin{eqnarray*}
\fl
  \VEV{\hat\sfF_{\sff}\hat\sfF_{\delta u_{(3)}}}=\Aip{\delta\sff^*}{\delta\delta u_{(3)}}_+=0
\hspace{11.15em} \Rightarrow \ \hat\sfF_{\delta u_{(3)}}\equiv 0
\ \Rightarrow \ \Intd\hat\sfF\equiv 0;\\
\fl
  \VEV{\hat\sfF_{\sff}\hat\sfF_{\Intd u_{(1)}}}=\Aip{\delta\sff^*}{\delta\Intd u_{(1)}}_+
     =\Aip{\delta\sff^*}{(\delta\Intd+\Intd\delta) u_{(1)}}_+=0
\ \Rightarrow \ \hat\sfF_{\Intd u_{(1)}}\equiv 0
\ \Rightarrow \ \delta\hat\sfF\equiv 0\\
\fl
  \quad\mbox{(because $\Aip{\delta\sff^*}{\Intd v}_\pm=0\ \forall v$, and for the other term because of projection to the light-cone);}\\
\fl\mbox{and }
\VEV{\hat\sfF_{\sff}\hat\sfF_{\sfg}}=\VEV{\hat\sfF_{\sff}\hat\sfF_{\sfg^{(+)}}}
\ \Rightarrow\ \hat\sfF_{\sfg}\equiv\hat\sfF_{\sfg^{(+)}},
\ \mbox{where }\widetilde{g^{(\pm)}}(k)=\tilde g(k)\theta(\pm k),
\end{eqnarray*}
so we can write $\hat\mathsf{A}$ as a potential for $\hat\sfF$, $\hat\sfF=\Intd\hat\mathsf{A}$, $\hat\sfF_\sff=\hat\mathsf{A}_{\delta\sff}$, and
we have $\VEV{\hat\mathsf{A}_u\hat\mathsf{A}_v}=\Aip{u^*}{v}_+$, albeit problematically because $\Aip{u}{v}_+$ is not a pre--inner product.

For a random field operator $\bbFF_{\sff}$, we have, using $P_\pm=\Half(1\pm\rmi\star)$ for brevity and clarity,
$$\VEV{\bbFF_{\sff}\bbFF_{\sfg}}
  =\Aip{\delta P_+\sff^*}{\delta P_+\sfg}_++\Aip{\delta P_-\sff^*}{\delta P_-\sfg}_-,$$
then, using $\delta P_\pm\delta=\pm\Half\rmi\delta\star\delta=\pm\Half\rmi\delta\Intd\star$ (when acting on 3--forms)
and $\delta P_\pm\Intd=\Half\delta\Intd$,
\begin{eqnarray*}\fl
  \VEV{\bbFF_{\sff}\bbFF_{\delta u_{(3)}}}=
{{\scriptstyle\frac{\scriptstyle\rmi}{\scriptstyle 2}}}\Aip{\delta P_+\sff^*}{\delta\Intd\star u_{(3)}}_+
    -{{\scriptstyle\frac{\scriptstyle\rmi}{\scriptstyle 2}}}\Aip{\delta P_-\sff^*}{\delta\Intd\star u_{(3)}}_- =0
                            \ \Rightarrow \ \Intd\bbFF\equiv 0;\\
\fl  \VEV{\bbFF_{\sff}\bbFF_{\Intd u_{(1)}}}=
\quad\Half\Aip{\delta P_+\sff^*}{\delta\Intd u_{(1)}}_+
    +\Half\Aip{\delta P_-\sff^*}{\delta\Intd u_{(1)}}_- =0
                            \hspace{1.2em} \Rightarrow \ \delta\bbFF\equiv 0;
\end{eqnarray*}
so $\bbFF$ still satisfies the Maxwell equations, however the projection to positive frequency becomes
$$\VEV{\bbFF_{\sff}\bbFF_{\sfg}}=\VEV{\bbFF_{\sff}\bbFF_{P_+\sfg^{(+)}+P_-\sfg^{(-)}}}
\qquad\Rightarrow\qquad\bbFF_{\sfg}\equiv\bbFF_{P_+\sfg^{(+)}+P_-\sfg^{(-)}}.$$
Because of the Hodge dual in this construction, to construct a potential for $\bbFF$ we have to introduce both a 1--form $\hat\mathbb{A}$ and a 3--form $\hat\mathbb{B}$, setting $\bbFF=\Intd\hat\mathbb{A}+\delta\hat\mathbb{B}$ so that
$\bbFF_\sff=\hat\mathbb{A}_{\delta\sff}+\hat\mathbb{B}_{\Intd\sff}$.
Defining $\hat\mathbb{X}_{u_{(1)}\oplus u_{(3)}}=\hat\mathbb{A}_{u_{(1)}}+\hat\mathbb{B}_{u_{(3)}}$,
we have $\bbFF_\sff=\hat\mathbb{X}_{\delta\sff\oplus\Intd\sff}$.
If we write
\begin{eqnarray*}\fl
\VEV{\hat\mathbb{X}_{u_{(1)}\oplus u_{(3)}}\hat\mathbb{X}_{v_{(1)}\oplus v_{(3)}}}\\
\fl\quad  =\Aip{\Half(u^*_{(1)}-\rmi{\star} u^*_{(3)})}{\Half(v_{(1)}-\rmi{\star} v_{(3)})}_+
   +\Aip{\Half(u^*_{(1)}+\rmi{\star} u^*_{(3)})}{\Half(v_{(1)}+\rmi{\star} v_{(3)})}_-,
\end{eqnarray*}
then, using that $\Intd\star=-{\star}\delta$ when acting on 2--forms,
\begin{eqnarray*}\fl
\VEV{\bbFF_{\sff}\bbFF_{\sfg}}&=
     \Aip{\Half(\delta\sff^*-\rmi{\star} \Intd\sff^*)}{\Half(\delta\sff^*-\rmi{\star} \Intd\sff^*)}_+
   +\Aip{\Half(\delta\sff^*+\rmi{\star} \Intd\sff^*)}{\Half(\delta\sff^*+\rmi{\star} \Intd\sff^*)}_-\\
\fl&=\Aip{\Half(\delta\sff^*+\rmi\delta{\star}\sff^*)}{\Half(\delta\sff^*+\rmi\delta{\star}\sff^*)}_+
   +\Aip{\Half(\delta\sff^*-\rmi\delta{\star}\sff^*)}{\Half(\delta\sff^*-\rmi\delta{\star}\sff^*)}_-\\
\fl&=\Aip{\delta P_+\sff^*}{\delta P_+\sfg}_++\Aip{\delta P_-\sff^*}{\delta P_-\sfg}_-.
\end{eqnarray*}

We could, for the quantized electromagnetic field, have set
$\hat\sfF=\Intd\hat\mathsf{A}+\delta\hat\mathsf{B}$, so that
$\hat\sfF=\hat\mathsf{A}_{\delta\sff}+\hat\mathsf{B}_{\Intd\sff}
  =\hat\mathsf{X}_{\delta\sff\oplus\Intd\sff}$, in which case to ensure that
$\VEV{\hat\sfF^\dagger_{\sff}\hat\sfF_{\sfg}}=\Aip{\delta\sff}{\delta\sfg}_+$ we would have to set
$\VEV{\hat\mathsf{X}_{u_{(1)}\oplus u_{(3)}}\hat\mathsf{X}_{v_{(1)}\oplus v_{(3)}}}
  =\Aip{u^*_{(1)}}{v_{(1)}}_+$, which has the effect that $\hat\mathsf{B}\equiv 0$.
The constructions of potentials for $\hat\sfF$ and $\bbFF$, as for the constructions in Section~\ref{EMfield}, have the same number of effective degrees of freedom, however with projections to different linear subspaces.

\section{Calculating fermionic VEVs\label{CalculateVEVs}}
It is worth showing briefly how the calculation of fermionic VEVs proceeds efficiently.
We first choose (part of) an orthogonal basis, then manipulate objects that are either orthogonal or parallel.
For $\VEV{\hat\psi^{\dagger}_{W}\hat\psi^{\ }_{V}
                                              \hat\psi^{\dagger}_{A}\hat\psi^{\ }_{B}
                                              \hat\psi^{\dagger}_{V}\hat\psi^{\ }_{W}}$, with $V$ and $W$ orthogonal, we choose $V$ and $W$ as a partial basis (if they were not orthogonal, we would first Gram-Schmidt orthogonalize), then write $A=A{\perp}V{\perp}W+A\|V+A\|W$, where $A\|V=A-A{\perp}V$ is the component of $A$ parallel to $V$, and similarly for $B$. Only three and then two terms of the nine terms in the expansion of $ \hat\psi^{\dagger}_{A}\hat\psi^{\ }_{B}$ survive,
\begin{eqnarray*}\fl
\VEV{\hat\psi^{\dagger}_{W}\hat\psi^{\ }_{V}
                            \hat\psi^{\dagger}_{A}\hat\psi^{\ }_{B}
                            \hat\psi^{\dagger}_{V}\hat\psi^{\ }_{W}}
       &=&\VEV{\hat\psi^{\dagger}_{W}\hat\psi^{\ }_{V}
                            \hat\psi^{\dagger}_{A{\perp}V{\perp}W}\hat\psi^{\ }_{B{\perp}V{\perp}W}
                            \hat\psi^{\dagger}_{V}\hat\psi^{\ }_{W}}\\
    &&\quad+\VEV{\hat\psi^{\dagger}_{W}\hat\psi^{\ }_{V}
                            \hat\psi^{\dagger}_{A\|V}\,\hat\psi^{\ }_{B{\|}V}\,
                            \hat\psi^{\dagger}_{V}\hat\psi^{\ }_{W}}\\
    &&\quad+\VEV{\hat\psi^{\dagger}_{W}\hat\psi^{\ }_{V}
                            \hat\psi^{\dagger}_{A\|W}\hat\psi^{\ }_{B{\|}W}
                            \hat\psi^{\dagger}_{V}\hat\psi^{\ }_{W}}\\
    &&\hspace{-6em}=\left[\strut(A{\perp}V{\perp}W,B{\perp}V{\perp}W)_+ +(A\|V,B\|V)\right]
           \VEV{\hat\psi^{\dagger}_{W}\hat\psi^{\ }_{V}\hat\psi^{\dagger}_{V}\hat\psi^{\ }_{W}},
\end{eqnarray*}
where the $\hat\psi^{\dagger}_{A\|W}\hat\psi^{\ }_{B{\|}W}$ term vanishes because $\{\hat\psi^{\ }_{V},\hat\psi^{\dagger}_{A\|W}\}=0$ and $\hat\psi^{\dagger}_{W}\hat\psi^{\dagger}_{A\|W}=0$. Note especially that factors move outside the vacuum state evaluation with or without a projection to positive frequency, $(\cdot,\cdot)_+$ or $(\cdot,\cdot)$, depending respectively on whether they are orthogonal to all other factors or a linear multiple of some other factor.
The two pre--inner products on (the infinite--dimensional) test function space are, as always, crucial.


\section*{References}
\begin{thebibliography}{00}
\bibitem{Koopman}
  Koopman B O 1931 \textit{Proc. Nat. Acad. Sci.} \textbf{17} 315

\bibitem{MorganEPL}
  Morgan P 2009 \textit{EPL} \textbf{87} 31002

\bibitem{Cohn}
  Cohn J 1980 \textit{Am. J. Phys.} \textbf{48} 379

\bibitem{MorganEMunpublished}
  Morgan P 2009 arXiv:0908.2439

\bibitem{TsangCaves}
  Tsang M and Caves C 2012 \textit{Phys. Rev.} X \textbf{2} 031016

\bibitem{Ghose}
   Rajagopal A K and Ghose P 2016 \textit{Pramana -- J. Phys.} \textbf{86} 1161

\bibitem{vonNeumann}
  von Neumann J 1932 \textit{Ann. Math.} \textbf{33} 587

\bibitem{Blum}
  Blum A S 2017 \textit{Studies in History and Philosophy of Modern Physics} \textbf{60} 46

\bibitem{BaezBiamonte}
  Baez J C and Biamonte J D 2018 \textit{Quantum Techniques In Stochastic Mechanics} (World Scientific Press, Singapore)

\bibitem{MumfordDesolneux}
  Mumford D and Desolneux A 2010 \textit{Pattern Theory: The Stochastic Analysis of Real-World Signals} (A K Peter, Natick, MA)
  
\bibitem{Rozanov}
  Rozanov Yu A 1998 \textit{Random Fields and Stochastic Partial Differential Equations} (Springer, Dordrecht)

\bibitem{MenikoffSharp}
  Menikoff R and Sharp D H 1977 \textit{J. Math. Phys.} \textbf{18} 471

\bibitem{Zurek}
  Zurek W H 2017 \textit{Phys. Rev.} A \textbf{76} 052110

\bibitem{Roberts}
  Roberts B 2018 \textit{Studies in History and Philosophy of Modern Physics} \textbf{63} 150 (arXiv:1610.07637)

\bibitem{Cohen}
  Cohen L 1989 \textit{Proc. IEEE} \textbf{77} 941

\bibitem{SW}
  Streater R F and Wightman A S 1964 \textit{PCT, spin \& statistics, and all that} (W. A. Benjamin, New York)

\bibitem{Haag}
  Haag R 1996 \textit{Local Quantum Physics} 2nd Edition (Springer, Berlin)

\bibitem{Neeb}
  Neeb K-H and \'Olafsson G 2018 arXiv:1802.09037

\bibitem{tHooft}
  't Hooft G 2016 \textit{The Cellular Automaton Interpretation of Quantum Mechanics} (Springer, Heidelberg)

\bibitem{JordanWigner}
  Jordan P and Wigner E 1928  \textit{Z. Phys.} \textbf{47} 631

\bibitem{IZ}
  Itzykson and Zuber J-B 1980 \textit{Quantum Field Theory} (McGraw-Hill, New York)

\bibitem{BoyerA}
  Boyer T H 1975 \textit{Phys. Rev. D} \textbf{11} 809

\bibitem{BoyerB}
  Boyer T H 1975 \textit{Phys. Rev. D} \textbf{11} 790

\bibitem{SED}
  de la Pe\~na L and Cetto A M 1996 \textit{The Quantum Dice: An Introduction to Stochastic Electrodynamics} (Kluwer, Dordrecht)

\bibitem{MorganBellRF}
  Morgan P 2006 \textit{J. Phys. A} \textbf{39} 7441

\bibitem{SegalGoodman}
  Segal I E and Goodman R W 1965 \textit{J. Math. Mech.} \textbf{14} 629

\bibitem{Hegerfeldt}
  Hegerfeldt G C 1998 \textit{Irreversibility and Causality: Semigroups and Rigged Hilbert Spaces} Eds. A Bohm, H-D Doebner, and P Kielanowski (Springer, Berlin) 238

\bibitem{Landau}
  Landau L J 1987 \textit{Phys. Lett. A} \textbf{120} 54

\bibitem{MorganTF}
  Morgan P 2015 arXiv:1507.08299

\bibitem{Rejzner}
  Rejzner K 2016 \textit{Perturbative Algebraic Quantum Field Theory} (Springer, Heidelberg)

\bibitem{Woodhouse}
  Woodhouse N M J 1991 \textit{Geometric Quantization} 2nd Edition (Oxford University Press, Oxford)

\bibitem{Zalamea}
  Zalamea F 2018 \textit{Found. Phys.} \textbf{48} 1061 (arXiv:1711.06914)

\bibitem{Kisil}
  Kisil V 2017 \textit{Geometry, Integrability and Quantization} \textbf{18} 11 (arXiv:1611.05650)

\bibitem{Mauro}
  Mauro D 2003 \textit{Phys. Lett.} A \textbf{315} 28

\bibitem{deGossonA}
  de Gosson M 2006 \textit{Symplectic Geometry and Quantum Mechanics} (Birkh\"auser, Basel)

\bibitem{deGossonB}
  de Gosson M A and Hiley B J 2001 \textit{Found. Phys.} \textbf{41} 1415

\bibitem{Blasone}
  Blasone M, Jizba P and Scardigli F 2009 \textit{J.Phys:Conf.Ser.} \textbf{174} 012034

\bibitem{Klein}
  Klein U 2018 \textit{Quantum Studies: Math. Found.} \textbf{5} 219

\bibitem{Rajagopal}
  Rajagopal A K and Ghose P 2016 \textit{Pramana - J. Phys.} \textbf{86} 1161

\bibitem{Viennot}
  Viennot D and Aubourg L 2018 \textit{J. Phys.} A \textbf{51} 335102


\end{thebibliography}
\end{document}